\documentclass[10pt,twocolumn,aps,prd,nofootinbib,superscriptaddress,floatfix]{revtex4-1}
\usepackage[utf8]{inputenc}
\usepackage{color}
\usepackage{amsmath}
\usepackage{amssymb}
\usepackage{graphicx}
\usepackage{subcaption}
\usepackage{verbatim}
\usepackage{comment}
\usepackage{wasysym}
\usepackage{textcomp}
\usepackage{amsfonts}
\usepackage{graphics}
\usepackage{epstopdf}
\usepackage{slashed}
\usepackage{multirow}
\usepackage{adjustbox}
\usepackage{lipsum}
\usepackage{rotating}
\usepackage{xcolor}
\usepackage{wrapfig}
\usepackage{epsfig}

\definecolor{blu}{rgb}{0, 0, 0.699219}
\definecolor{bluc}{rgb}{0, 0, 0.898438}
\definecolor{blus}{rgb}{0, 0, 0.398438}
\definecolor{rosso}{rgb}{0.601562, 0, 0}
\definecolor{rossoc}{rgb}{0.800781, 0, 0}
\definecolor{rossos}{rgb}{0.449219, 0, 0}
\definecolor{verde}{rgb}{0.0585938, 0.75, 0.308594}
\definecolor{verdec}{rgb}{0.0664062, 0.851562, 0.347656}
\definecolor{verdes}{rgb}{0.046875, 0.601562, 0.246094}
\usepackage[bookmarks=false,breaklinks=true,pdfborder={0 0 1},backref=false,colorlinks=false]
{hyperref}
\hypersetup{pdfstartview=FitH}
\usepackage[compat=1.1.0]{tikz-feynman}
\DeclareRobustCommand{\orcidicon}{\hspace{-2.1mm}
\begin{tikzpicture}
\draw[lime,fill=lime] (0,0.0) circle [radius=0.13] node[white] {{\sffamily\tiny ID}}; \draw[white,fill=white] (-0.0525,0.095) circle [radius=0.007];
\end{tikzpicture} \hspace{-3.7mm}}
\foreach \x in {A, ..., Z} {\expandafter\xdef\csname orcid\x\endcsname{\noexpand\href{https://orcid.org/\csname orcidauthor\x\endcsname} {\noexpand\orcidicon}}}

\begin{document}
\title{Neutrino Masses, Dark Matter, and Collider Constraints in a $\mathbf{[SU(2)\times U(1)]^{2}}$ Model}
\author{Amine~Ahriche\,\,\orcidA}
\email{ahriche@sharjah.ac.ae}

\affiliation{Department of Applied Physics and Astronomy, University of Sharjah,
P.O. Box 27272 Sharjah, UAE.}
\affiliation{Research Institute of Sciences and Engineering, University of Sharjah,
P.O. Box 27272 Sharjah, UAE.}
\author{Nabil~Baouche\,\,\orcidN}
\email{baouche.nabil@gmail.com}

\affiliation{Faculty of Science and Technology, University of Jijel PB 98 Ouled
Aissa, DZ-18000 Jijel, Algeria.}
\affiliation{Laboratoire de Physique des Particules et Physique Statistique, Ecole
Normale Superieure, BP 92 Vieux Kouba, DZ-16050 Algiers, Algeria.}
\author{Aldo~Deandrea\,\,\orcidD}
\email{deandrea@ip2i.in2p3.fr}

\affiliation{Université Lyon 1, CNRS, IP2I, UMR 5822, Villeurbanne, France.}
\affiliation{Department of Physics, University of Johannesburg, PO Box 524, Auckland
Park 2006, South Africa.}
\author{Mohammed~Omer~Khojali\,\,\orcidK}
\email{m.khojali@ip2i.in2p3.fr}

\affiliation{Université Lyon 1, CNRS, IP2I, UMR 5822, Villeurbanne, France.}

\author{Salah~Nasri\,\,\orcidK}
\email{snasri@uaeu.ac.ae, salah.nasri@cern.ch}

\affiliation{Physics Department, UAE University, POB 17551, Al Ain, United Arab
Emirates.}
\begin{abstract}
We study an extension of the Standard Model based on the gauge group
$SU(2)_{L}\times SU(2)_{H}\times U(1)_{1}\times U(1)_{2}$ that links
the origin of tiny neutrino masses to dark matter. The particle content
is enlarged by a hidden $SU(2)_{H}$ scalar doublet, a complex scalar
singlet, an inert $SU(2)_{L}$ scalar doublet, and three Majorana
singlet fermions. An exact $Z_{2}$ symmetry forbids the tree-level
neutrino Dirac mass, so that active neutrino masses are generated
at one loop through the scotogenic mechanism, and it stabilizes the
lightest $Z_{2}$-odd particle. Depending on the mass hierarchy, the
dark matter candidate is a Majorana fermion ($N_{1}$), an inert scalar
($H^{0}$), or, owing to a residual symmetry of the hidden sector,
a hidden vector boson ($Q^{0}$). Through a random parameter scan,
we impose perturbative unitarity, vacuum stability, electroweak precision
tests, LHC Higgs signal strengths and total width, LHC dilepton resonance
searches for the heavy neutral gauge bosons $Z^{\prime}$ and $Z^{\prime\prime}$, lepton
flavor violation limits, and the LZ direct detection bounds. We identify
parameter regions compatible with these constraints and discuss how
the extended gauge and scalar sectors shape the direct detection prospects
of each candidate. Relic abundance and indirect detection are discussed
only qualitatively. 
\end{abstract}
\maketitle

\section{Introduction}

The discovery of the Higgs boson at the Large Hadron Collider (LHC)
represented a major milestone in particle physics, completing the
experimental discovery of the Standard Model (SM) particle content~\cite{ATLAS:2012yve,CMS:2012qbp}.
However, two observations still require physics beyond its minimal
form: neutrino oscillations imply nonzero neutrino masses~\cite{Super-Kamiokande:2001ljr,Esteban:2020cvm},
and cosmological and astrophysical data indicate a non-baryonic dark
matter (DM) component~\cite{Bertone:2016nfn,Bertone:2004pz,Sofue:2000jx}.
The smallness of neutrino masses motivates scenarios in which the
leading contribution arises radiatively. Among these, the scotogenic
mechanism provides a framework in which neutrino mass generation is
linked to dark matter~\cite{Ma:2006km}. In this setup, an inert
scalar doublet and Majorana singlet fermions generate neutrino masses
at one loop, while an exact discrete symmetry forbids a tree-level
Dirac mass term and stabilizes the lightest odd particle. The same
new particles can therefore connect radiative neutrino mass generation to either a scalar or a fermionic DM candidate.

An enlarged hidden gauge sector provides a complementary possibility:
a neutral vector DM candidate together with heavy neutral gauge bosons
potentially accessible to collider searches through their interactions
with SM particles~\cite{Hambye:2008bq,Baouche:2021wwa,Gross:2015cwa}.
Combining a hidden gauge sector with radiative neutrino mass generation
is motivated because the symmetry breaking scale and scalar mixing
affect several observables at once. They determine the masses and
couplings of new vectors, modify Higgs properties, and control interactions
relevant to DM annihilation and detection. A viable spectrum must
therefore be assessed against precision measurements, collider searches, and neutrino data, and DM direct-detection bounds within the same
parameter space. For a neutral hidden vector to serve as a DM candidate,
its stability must be ensured by an exact residual symmetry that forbids
its decay into lighter states. In this model, $Q^{0}$ is even under
the imposed $Z_{2}$; its stability follows from the residual parity described in Sec.~\ref{Sec:DM}

In this work, we investigate the interplay between radiative neutrino
mass generation and dark matter phenomenology in an extension based
on $SU(2)_{L}\times SU(2)_{H}\times U(1)_{1}\times U(1)_{2}$, with
the usual color group understood. The additional fields comprise an
inert electroweak scalar doublet $\eta$, a hidden $SU(2)_{H}$ scalar
doublet $\Phi$, a complex scalar singlet $S$, and three singlet
fermions $N_{i}$. The central question is whether the extended particle
content can simultaneously accommodate radiative neutrino mass generation
and a viable dark matter candidate, consistent with the experimental
bounds considered here. To address the neutrino sector, we use the
Casas-Ibarra parametrization~\cite{Casas:2001sr} to relate the Yukawa
couplings to the neutrino masses and mixing parameters, and discuss
the associated charged lepton flavor violation bounds. We then examine
constraints from perturbative unitarity of high energy scalar scattering,
vacuum stability, electroweak oblique parameters, Higgs decay width
and signal strengths, and LHC dilepton searches for the heavy neutral
gauge bosons $Z^{\prime}$ and $Z^{\prime\prime}$~\cite{ATLAS:2019erb,CMS:2026xoc,Accomando:2010fz}.
For fermionic candidates $N_{1}$, inert scalar $H_{0}$, and hidden vector $Q^{0}$
, we evaluate the spin-independent DM
nuclear-scattering cross sections from scalar and gage boson exchange, as
applicable, and compare them with direct-detection limits. The numerical
results display parameter correlations and illustrate the compatibility
with these bounds. Annihilation channels and indirect detection prospects
are discussed qualitatively.

The paper is organized as follows. Section~\ref{Sec:Model} presents
the model and symmetry breaking, Sec.~\ref{Sec:Neutrino} describes
the neutrino mass generation mechanism and associated flavor constraints,
and Sec.~\ref{Sec:Constraints} collects the theoretical and experimental
bounds. Section~\ref{Sec:DM} discusses the DM candidates and their
interactions, Secs.~\ref{sec:DM_DD} and \ref{sec:DM_ID} address
direct and indirect detection, respectively, while Sec.~\ref{Sec:Numeric}
presents the numerical results. We conclude our paper in Sec.~\ref{Sec:Conclusion}.

\section{The Model~\label{Sec:Model}}

We consider the gauge symmetry group $SU(2)_{L}\times SU(2)_{H}\times U(1)_{1}\times U(1)_{2}$,
with the usual QCD factor $SU(3)_{c}$ understood. In addition to
the SM fields, the model contains an inert electroweak scalar doublet
($\eta$), an $SU(2)_{H}$ scalar doublet ($\Phi$), a complex scalar
singlet ($S$) and three singlet fermions ($N_{1,2,3}$). An exact
$Z_{2}$ symmetry is imposed, under which $\eta$ and $N_{i}$ are
odd and all other fields are even. All SM fields are singlets under
$SU(2)_{H}$, while they may carry charges under the two abelian gauge
symmetries. The SM hypercharge is embedded as $Y=n_{1}+n_{2}$, where
$n_{1,2}$ denotes the charges under $U(1)_{1,2}$. We denote the quantum
numbers by $(d_{c},d_{L},d_{H},n_{1},n_{2})$, where $d_{c}$, $d_{L}$,
and $d_{H}$ are the dimensions of the representations under $SU(3)_{c}$,
$SU(2)_{L}$, and $SU(2)_{H}$, respectively. We use the anomaly-free
charge assignment derived in Appendix~\ref{app:anom}: 
\begin{align*}
\Phi\sim(1,1,2,\tfrac{1}{4},-\tfrac{1}{4}),\quad S & \sim(1,1,1,-1,1),\\
H\sim(1,2,1,\tfrac{1}{4},\tfrac{1}{4}),\quad\eta & \sim(1,2,1,\tfrac{1}{4},\tfrac{1}{4}),\\
Q_{L}\sim(3,2,1, & -\tfrac{1}{12},\tfrac{1}{4}),\\
u_{R}\sim(3,1,1,\tfrac{1}{6},\tfrac{1}{2}),\quad & d_{R}\sim(3,1,1,-\tfrac{1}{3},0),\\
L_{L}\sim(1,2,1, & \tfrac{1}{4},-\tfrac{3}{4}),\\
\ell_{R}\sim(1,1,1,0,-1),\quad & N_{1,2,3}\sim(1,1,1,\tfrac{1}{2},-\tfrac{1}{2}).
\end{align*}
where the quantum numbers of the particles are indicated with a 5-component
vector corresponding to the $(SU(3),SU(2)_{L},SU(2)_{H},U(1)_{1},U(1)_{2})$
multiplets. The Lagrangian can be expressed as 
\begin{equation}
\mathcal{L}=\mathcal{L}_{\text{kinetic}}+\mathcal{L}_{\text{Yukawa}}-V,
\end{equation}
where $V$ denotes the scalar potential. The kinetic terms and gauge
interactions are contained in 
\begin{equation}
\begin{aligned}\mathcal{L}_{kin}={} & \sum_{\psi}\bar{\psi}i\gamma^{\mu}D_{\mu}\psi+\sum_{\phi}(D^{\mu}\phi)^{\dagger}(D_{\mu}\phi)\\
 & -\frac{1}{4}W^{a\mu\nu}W^{a}_{\mu\nu}-\frac{1}{4}Q^{c\mu\nu}Q^{c}_{\mu\nu}\\
 & -\frac{1}{4}B^{1\mu\nu}B^{1}_{\mu\nu}-\frac{1}{4}B^{2\mu\nu}B^{2}_{\mu\nu}.
\end{aligned}
\end{equation}
with the $a,c=1,2,3$, $\psi$ and $\phi$ running over fermion and
scalar multiplets, respectively. The covariant derivative and the
corresponding field strength tensors are given by 
\begin{align}
D_{\mu} & =\partial_{\mu}+igW^{b}_{\mu}\frac{\tau_{b}}{2}+ig_{H}Q^{c}_{\mu}\frac{\tau_{c}}{2}+ig_{1}'n_{1}B^{1}_{\mu}+ig_{2}'n_{2}B^{2}_{\mu},\nonumber \\
W^{b}_{\mu\nu} & =\partial_{\mu}W^{b}_{\nu}-\partial_{\nu}W^{b}_{\mu}+g\epsilon^{abc}W^{a}_{\mu}W^{c}_{\nu},\nonumber \\
Q^{c}_{\mu\nu} & =\partial_{\mu}Q^{c}_{\nu}-\partial_{\nu}Q^{c}_{\mu}+g_{H}\epsilon^{cab}Q^{a}_{\mu}Q^{b}_{\nu},\nonumber \\
B^{1}_{\mu\nu} & =\partial_{\mu}B^{1}_{\nu}-\partial_{\nu}B^{1}_{\mu},\quad B^{2}_{\mu\nu}=\partial_{\mu}B^{2}_{\nu}-\partial_{\nu}B^{2}_{\mu}.\label{eq:Cov}
\end{align}
Here, $(g,g_{H},g'_{1},g'_{2})$ and $(W^{a}_{\mu},Q^{a}_{\mu},B^{1}_{\mu},B^{2}_{\mu})$
denote the gauge couplings and gauge fields associated with $SU(2)_{L}\times SU(2)_{H}\times U(1)_{1}\times U(1)_{2}$,
respectively, while $\tau_{a}$ ($a=1,2,3$) are the Pauli matrices.
For the Abelian gauge couplings, we adopt the symmetric choice $g'_{1}=g'_{2}=\sqrt{2}\,g_{1}$,
where $g_{1}\equiv g_{Y}$ denotes the SM hypercharge gauge coupling.
The Yukawa Lagrangian is given by 
\begin{equation}
\begin{aligned}\mathcal{L}\supset{} & -y_{L}\overline{L}_{L}He_{R}-y_{d}\overline{Q}_{L}Hd_{R}\\
 & -y_{u}\overline{Q}_{L}\tilde{H}u_{R}-f_{i\alpha}\overline{L}_{\alpha L}\tilde{\eta}N_{i}\\
 & -\frac{1}{2}y_{N_{i}}S(N^{T}_{i})^{C}N_{i}+\text{h.c.}
\end{aligned}
\label{eq:Yuk:1}
\end{equation}
where $\tilde{H}=i\sigma_{2}H^{*}$ and $\tilde{\eta}=i\sigma_{2}\eta^{*}$.
The field content and charge assignments cancel the gauge and mixed
gravitational anomalies, as detailed in Appendix~\ref{app:anom}.
\vspace{-0.6cm}

\subsection{Symmetry breaking and neutral gauge bosons}

The symmetry breaking pattern in the model proceeds in two stages,
\begin{equation}
\begin{aligned} & SU(3)_{C}\times SU(2)_{L}\times SU(2)_{H}\times U(1)_{1}\times U(1)_{2}\\
 & \quad\xrightarrow{\langle\Phi\rangle,\langle S\rangle}SU(3)_{C}\times SU(2)_{L}\times U(1)_{Y}\\
 & \quad\xrightarrow{\langle H\rangle}SU(3)_{C}\times U(1)_{\mathrm{em}}.
\end{aligned}
\end{equation}
In the first stage, the vacuum expectation values (VEVs) of $\Phi$
and $S$ break the extended gauge symmetry to that of the SM. With
the chosen Abelian gauge couplings, the hypercharge gauge field is
$B_{\mu}=(B^{1}_{\mu}+B^{2}_{\mu})/\sqrt{2}$, while the orthogonal
combination $(B^{1}_{\mu}-B^{2}_{\mu})/\sqrt{2}$ mixes with $Q^{3}_{\mu}$
to form the massive gauge bosons $Z^{\prime}$ and $Z^{\prime\prime}$. The second stage
corresponds to the usual electroweak symmetry breaking induced by
the SM Higgs doublet. Neglecting Abelian gauge kinetic mixing, the
neutral gauge eigenstates are related to the physical states by 
\begin{equation}
\begin{aligned}A_{\mu} & =s_{W}W^{3}_{\mu}+\frac{c_{W}}{\sqrt{2}}(B^{1}_{\mu}+B^{2}_{\mu}),\\
Z_{\mu} & =c_{W}W^{3}_{\mu}-\frac{s_{W}}{\sqrt{2}}(B^{1}_{\mu}+B^{2}_{\mu}),\\
Z^{\prime}_{\mu} & =c_{\alpha}Q^{3}_{\mu}-\frac{s_{\alpha}}{\sqrt{2}}(B^{1}_{\mu}-B^{2}_{\mu}),\\
Z^{\prime\prime}_{\mu} & =s_{\alpha}Q^{3}_{\mu}+\frac{c_{\alpha}}{\sqrt{2}}(B^{1}_{\mu}-B^{2}_{\mu}).
\end{aligned}
\label{eq:eigen}
\end{equation}
Here $s_{W}=\sin\theta_{W}$, $c_{W}=\cos\theta_{W}$, with $\theta_{W}$
denoting the Weinberg angle, while $s_{\alpha}=\sin\alpha$ and $c_{\alpha}=\cos\alpha$,
where $\alpha$ is the mixing angle in the $\{Q^{3}_{\mu},(B^{1}_{\mu}-B^{2}_{\mu})/\sqrt{2}\}$
sector. The $W^{\pm}$ gauge bosons have the same definition as in
the SM, and in addition we have two mass degenerate conjugate vector
states, $Q^{0}$ and $\bar{Q}^{0}$, with $Q^{0}_{\mu}=(Q^{1}_{\mu}+iQ^{2}_{\mu})/\sqrt{2}$.
The gauge field $Q^{0}$ does not couple to fermions, while $Z^{\prime}$
and $Z^{\prime\prime}$ couple to fermions as 
\begin{equation}
\begin{aligned}\mathcal{L} & \supset\frac{e}{c_{W}}J^{\mu}\bigl(-s_{\alpha}Z^{\prime}_{\mu}+c_{\alpha}Z^{\prime\prime}_{\mu}\bigr),\\
J^{\mu} & =-\tfrac{1}{3}\bar{u}\gamma^{\mu}u-\tfrac{1}{3}\bar{d}\gamma^{\mu}d+\bar{e}\gamma^{\mu}e\\
 & \quad+\bar{\nu}_{e}\gamma^{\mu}P_{L}\nu_{e}+\bar{N}_{i}\gamma^{\mu}P_{R}N_{i}.
\end{aligned}
\end{equation}
where $P_{L}$ and $P_{R}$ denote the left- and right-handed chiral
projectors, respectively, with summation over fermion generations
understood. Among the new gauge bosons, only $Q^{0}$ can serve as
a DM candidate, while $Z^{\prime}$ and $Z^{\prime\prime}$ are unstable.
\vspace{-0.6cm}

\subsection{Scalar sector}

The scalar fields are parameterized as 
\begin{align}
H=\left(\begin{array}{c}
\chi^{+}\\
\frac{1}{\sqrt{2}}(\upsilon+h_{1}+ia_{1})
\end{array}\right), & \,\,\,\,\,\,\,S=\frac{1}{\sqrt{2}}(\upsilon_{s}+h_{2}+ia_{2}),\nonumber \\
\Phi=\left(\begin{array}{c}
\phi^{0}\\
\frac{1}{\sqrt{2}}(\upsilon_{\Phi}+h_{3}+ia_{3})
\end{array}\right), & \,\,\,\,\,\,\,\eta=\left(\begin{array}{c}
H^{+}\\
\frac{1}{\sqrt{2}}(H_{0}+iA_{0})
\end{array}\right),\label{eq:U}
\end{align}
and the most general renormalizable scalar potential consistent with
symmetries is 

\begin{eqnarray}
 & V & (H,\Phi,S,\eta)=-\mu^{2}_{H}|H|^{2}-\mu^{2}_{\Phi}|\Phi|^{2}\nonumber \\
 & - & \mu^{2}_{S}|S|^{2}+\mu^{2}_{\eta}|\eta|^{2}+\tfrac{1}{6}\lambda_{H}|H|^{4}+\tfrac{1}{6}\lambda_{\Phi}|\Phi|^{4}\nonumber\\
 & + & \tfrac{1}{6}\lambda_{S}|S|^{4}+\tfrac{1}{6}\lambda_{\eta}|\eta|^{4}+\lambda_{3}|H|^{2}|\eta|^{2}+\lambda_{4}|H^{\dagger}\eta|^{2}\nonumber \\
 & + & \omega_{1}|H|^{2}|\Phi|^{2}+\omega_{2}|H|^{2}|S|^{2}+\omega_{3}|\Phi|^{2}|\eta|^{2}+\omega_{4}|\Phi|^{2}|S|^{2}\nonumber \\
 & + & \omega_{5}|\eta|^{2}|S|^{2}+\tfrac{1}{2}\bigl\{\lambda_{5}(H^{\dagger}\eta)^{2}+\text{H.c.}\bigr\}.
 \label{eq:V}
\end{eqnarray}
The tree level tadpole conditions are 
\begin{equation}
\mu^{2}_{i}=\frac{1}{6}\Big(\lambda_{i}\upsilon^{2}_{i}+3\sum_{j\neq i}\omega_{ij}\upsilon^{2}_{j}\Big),\qquad i,j=H,\Phi,S,
\label{eq:Tree1}
\end{equation}
with $\upsilon_{H}=\upsilon$, and $\omega_{H\Phi}=\omega_{1}$, $\omega_{HS}=\omega_{2}$,
and $\omega_{\Phi S}=\omega_{4}$. The inert scalar $H_{0}$ does
not mix with the other CP-even states because of the unbroken $Z_{2}$
symmetry, and has a mass $m^{2}_{H_{0}}=\mu^{2}_{\eta}+\frac{1}{2}(\lambda_{+}\upsilon^{2}+\omega_{3}\upsilon^{2}_{\Phi}+\omega_{5}\upsilon^{2}_{s})$.
In the basis $(h_{1},h_{2},h_{3})$, the remaining CP-even mass matrix
is 
\begin{equation}
M^{2}_{\text{even}}=\begin{pmatrix}\frac{1}{3}\lambda_{H}\upsilon^{2} & \omega_{2}\upsilon\upsilon_{s} & \omega_{1}\upsilon\upsilon_{\Phi}\\
\omega_{2}\upsilon\upsilon_{s} & \frac{1}{3}\lambda_{S}\upsilon^{2}_{s} & \omega_{4}\upsilon_{\Phi}\upsilon_{s}\\
\omega_{1}\upsilon\upsilon_{\Phi} & \omega_{4}\upsilon_{\Phi}\upsilon_{s} & \frac{1}{3}\lambda_{\Phi}\upsilon^{2}_{\Phi}
\end{pmatrix},
\end{equation}
with $\lambda_{\pm}=\lambda_{3}+\lambda_{4}\pm\lambda_{5}$ . This
matrix is diagonalized using a $3\times3$ rotation matrix $R_{3\times3}$
yielding the mass eigenvalues $m^{2}_{H_{1}},m^{2}_{H_{2}}$ and $m^{2}_{H_{3}}$,
where the SM-like Higgs can be identified as any of the eigenstates
$H_{i}$. Here, we consider the case where $H_{1}$ serves as the
SM-like Higgs and $H_{2},H_{3}$ correspond to the heavier Higgs states.

In our setup, we have one physical CP-odd scalar ($A_{0}$) and one
pair of charged ($H^{\pm}$) scalars, whose mass squared matrices
are given by 
\begin{equation}
\begin{aligned}M^{2}_{\mathrm{odd}} & =\operatorname{diag}\{0,0,0,m^{2}_{A_{0}}\},\\
m^{2}_{A_{0}} & =\mu^{2}_{\eta}+\tfrac{1}{2}\bigl(\lambda_{-}\upsilon^{2}+\omega_{3}\upsilon^{2}_{\Phi}+\omega_{5}\upsilon^{2}_{s}\bigr),\\
M^{2}_{\mathrm{ch}} & =\operatorname{diag}\{0,m^{2}_{H^{\pm}}\},\\
m^{2}_{H^{\pm}} & =\mu^{2}_{\eta}+\tfrac{1}{2}\bigl(\lambda_{3}\upsilon^{2}+\omega_{3}\upsilon^{2}_{\Phi}+\omega_{5}\upsilon^{2}_{s}\bigr),
\end{aligned}
\end{equation}
in the bases $\{a_{1},a_{2},a_{3},A_{0}\}$ and $\{\chi^{+},H^{+}\}$,
respectively. The vanishing eigenvalues correspond to the neutral
and charged Goldstone bosons $(a_{1},a_{2},a_{3},\chi^{\pm})$, while
the non zero eigenvalues $m^{2}_{A_{0}}$ and $m^{2}_{H^{\pm}}$ correspond
to the CP odd and charged scalars, respectively. the complex scalar
field $\phi^{0}$ provides the Goldstone modes eaten by $Q^{0}$ and
$\bar{Q}^{0}$.

For the numerical analysis, we trade the quartic couplings for physical
masses, VEVs, and mixing parameters. The resulting relations are 
\begin{equation}
\begin{aligned}\lambda_{H} & =\frac{3}{\upsilon^{2}}\bigl(m^{2}_{H_{1}}R^{2}_{11}+m^{2}_{H_{2}}R^{2}_{21}+m^{2}_{H_{3}}R^{2}_{31}\bigr),\\
\lambda_{\Phi} & =\frac{3}{\upsilon^{2}_{\Phi}}\bigl(m^{2}_{H_{1}}R^{2}_{13}+m^{2}_{H_{2}}R^{2}_{23}+m^{2}_{H_{3}}R^{2}_{33}\bigr),\\
\lambda_{S} & =\frac{3}{\upsilon^{2}_{s}}\bigl(m^{2}_{H_{1}}R^{2}_{12}+m^{2}_{H_{2}}R^{2}_{22}+m^{2}_{H_{3}}R^{2}_{32}\bigr),\\
\lambda_{3} & =\frac{1}{\upsilon^{2}}\bigl(2m^{2}_{H^{\pm}}-\omega_{3}\upsilon^{2}_{\Phi}-\omega_{5}\upsilon^{2}_{s}-2\mu^{2}_{\eta}\bigr),\\
\lambda_{4} & =\frac{1}{\upsilon^{2}}\bigl(m^{2}_{H_{0}}+m^{2}_{A_{0}}-2m^{2}_{H^{\pm}}\bigr),\\
\lambda_{5} & =\frac{1}{\upsilon^{2}}\big(m^{2}_{H_{0}}-m^{2}_{A_{0}}\big),\\
\omega_{1} & =\frac{1}{\upsilon\upsilon_{\Phi}}\big(m^{2}_{H_{1}}R_{11}R_{13}+m^{2}_{H_{2}}R_{21}R_{23}+m^{2}_{H_{3}}R_{31}R_{33}\big),\\
\omega_{2} & =\frac{1}{\upsilon\upsilon_{s}}\big(m^{2}_{H_{1}}R_{11}R_{12}+m^{2}_{H_{2}}R_{21}R_{22}+m^{2}_{H_{3}}R_{31}R_{32}\big),\\
\omega_{4} & =\frac{1}{\upsilon_{\Phi}\upsilon_{s}}\big(m^{2}_{H_{1}}R_{12}R_{13}+m^{2}_{H_{2}}R_{22}R_{23}+m^{2}_{H_{3}}R_{32}R_{33}\big).
\end{aligned}
\end{equation}
The scalar self couplings and gauge scalar interactions relevant to
our analysis are given in Appendix~\ref{app:SV}.

\section{Neutrino Mass Generation~\label{Sec:Neutrino}}

\begin{figure}[h]
\begin{centering}
\includegraphics[width=0.95\columnwidth]{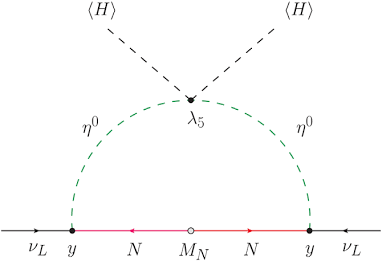} 
\par\end{centering}
\caption{Feynman diagram responsible for the neutrino mass.}
\label{fig:NuM} 
\end{figure}

The Yukawa interaction in Eq.~\eqref{eq:Yuk:1} couples lepton doublets
to the inert scalar doublet $\eta$ and to singlet fermions $N_{i}$.
The latter acquire Majorana masses $M_{i}=y_{N_{i}}\upsilon_{s}/\sqrt{2}$
once the singlet scalar $S$ develops a VEV. Due to the exact $Z_{2}$
symmetry, under which $\eta$ and $N_{i}$ are odd while all SM fields
are even, the Yukawa interaction $\overline{L_{L}}\tilde{H}N$ is
forbidden at tree level. Neutrino masses are therefore generated at
the one-loop level through the exchange of the inert scalar and Majorana
fermions $N_{i}$, as illustrated in Fig.~\ref{fig:NuM}. Writing
$\eta_{0}=(H_{0}+iA_{0})/\sqrt{2}$, the scalar potential induces
a mass splitting between the CP-even and CP-odd states $H_{0}$ and
$A_{0}$, enabling the generation of neutrino masses at one loop.
The resulting matrix elements are given by~\cite{Ma:2006km,Merle:2015gea}
\begin{equation}
\begin{aligned}(M_{\nu})_{\alpha\beta} & =\sum_{i}\frac{f_{i\alpha}f_{i\beta}M_{i}}{32\pi^{2}}\\
 & \quad\times\left[\mathcal{F}\left(\frac{m^{2}_{H_{0}}}{M^{2}_{i}}\right)-\mathcal{F}\left(\frac{m^{2}_{A_{0}}}{M^{2}_{i}}\right)\right].
\end{aligned}
\label{eq:num}
\end{equation}
Here, $\mathcal{F}(x)=x\log(x)/(x-1)$, $m^{2}_{H_{0}}$ and $m^{2}_{A_{0}}$
are the squared masses of the scalar fields $H_{0}$ and $A_{0}$,
respectively. 

Eq.~\eqref{eq:num} can be expressed compactly as 
\begin{equation}
M_{\nu}=f^{T}.\Lambda.f,\label{eq:numat}
\end{equation}
where 
\begin{equation}
\Lambda_{i}=\frac{M_{i}}{32\pi^{2}}\left[\mathcal{F}\!\left(\frac{m^{2}_{H_{0}}}{M^{2}_{i}}\right)-\mathcal{F}\!\left(\frac{m^{2}_{A_{0}}}{M^{2}_{i}}\right)\right].\label{eq:Lambda}
\end{equation}
We adopt the diagonalization convention 
\begin{equation}
U^{\dagger}_{\nu}M_{\nu}U^{*}_{\nu}=D_{m}\equiv\operatorname{diag}(m_{1},m_{2},m_{3}),\label{eq:nu_diagonalization}
\end{equation}
where $m_{i}\geq0$ are the light-neutrino mass eigenvalues and $U_{\nu}$
is the Pontecorvo--Maki--Nakagawa--Sakata (PMNS) mixing matrix~\cite{Pontecorvo:1967fh}.
For nonzero $\Lambda_{i}$, the Yukawa couplings can then be written
using the Casas--Ibarra parametrization~\cite{Casas:2001sr}, 
\begin{equation}
f=D_{\Lambda^{-1/2}}\,R\,D_{\sqrt{m}}\,U^{T}_{\nu},\label{eq:CasasIbarra}
\end{equation}
with 
\begin{align}
D_{\Lambda^{-1/2}} & =\operatorname{diag}\left(\Lambda^{-1/2}_{1},\Lambda^{-1/2}_{2},\Lambda^{-1/2}_{3}\right),\\
D_{\sqrt{m}} & =\operatorname{diag}\left(\sqrt{m_{1}},\sqrt{m_{2}},\sqrt{m_{3}}\right).
\end{align}
Here, $R$ is an arbitrary complex orthogonal matrix satisfying $R^{T}R=\mathbf{1}$.
For negative $\Lambda_{i}$, the inverse square root incorporates
the corresponding complex phase. Choosing $m_{i}$ and $U_{\nu}$
consistently with neutrino oscillation data~\cite{Tortola:2012te}
ensures that Eq.~\eqref{eq:CasasIbarra} reproduces the required
neutrino mass matrix.

For a characteristic Yukawa coupling $f$, the scale of an individual
contribution is $m^{{\rm typ}}_{\nu}\sim f^{2}|\Lambda_{k}|$. Thus,
\begin{align}
\frac{m^{{\rm typ}}_{\nu}}{0.05\,\mathrm{eV}} & \backsim\left(\frac{f}{0.01}\right)^{2}\left(\frac{M_{k}}{50\,\textrm{GeV}}\right)\left(\frac{\Lambda_{k}/M_{k}}{10^{-8}}\right).
\end{align}
This estimate assumes order unity flavor factors and no significant
cancellations among the contributions. The required suppression of
the loop coefficients $\Lambda_{k}$ arises from a small mass splitting
between the neutral inert scalars. Defining $\bar{m}^{2}=(m^{2}_{H_{0}}+m^{2}_{A_{0}})/2$,
and using $m^{2}_{H_{0}}-m^{2}_{A_{0}}=\lambda_{5}\upsilon^{2}$,
the nearly degenerate limit, $|\lambda_{5}|\upsilon^{2}\ll\bar{m}^{2}$,
gives 
\begin{equation}
\begin{aligned}\hspace{-0.15cm}\Lambda_{k}\simeq{}\frac{\lambda_{5}\upsilon^{2}}{32\pi^{2}}\frac{M_{k}}{\bar{m}^{2}-M^{2}_{k}}\times\left[1-\frac{M^{2}_{k}}{\bar{m}^{2}-M^{2}_{k}}\ln\!\left(\frac{\bar{m}^{2}}{M^{2}_{k}}\right)\right].\label{eq:Lambda_approx}\end{aligned}
\end{equation}
Consequently, the value of $|\lambda_{5}|$ compatible with both neutrino
data and the observed dark matter relic abundance depends on the Yukawa
couplings, the particle spectrum, and the flavor structure, together
with the dominant annihilation and co-annihilation processes. In parameter
regions where reproducing the neutrino masses requires $|\lambda_{5}|\ll1$,
this suppression is technically natural: in the minimal scotogenic
model, the limit $\lambda_{5}\to0$ restores a global lepton-number
symmetry~\cite{Merle:2015gea}. Additional singlet scalars~\cite{Ahriche:2020pwq}
or a second inert doublet~\cite{Ahriche:2022bpx, AbuSiam:2025voc} can modify the
loop structure, allowing neutrino masses and dark matter phenomenology
to be accommodated through a broader range of scalar-sector parameters.
The same Yukawa interactions also induce charged lepton flavor violating
(LFV) decays through one-loop diagrams involving $H^{\pm}$ and $N_{k}$
particles as in~\cite{Toma:2013zsa,Chekkal:2017eka}. Using the analytical
expressions in Ref.~\cite{Toma:2013zsa}, we evaluate $\ell^{-}_{\alpha}\to\ell^{-}_{\beta}\gamma$
and $\ell^{-}_{\alpha}\to\ell^{-}_{\beta}\ell^{+}_{\beta}\ell^{-}_{\beta}$
and require them to satisfy the corresponding experimental upper limits.

\section{Theoretical and Experimental Constraints~\label{Sec:Constraints}}

In the following, we discuss different theoretical and experimental
constraints that are relevant to our setup.

\subsubsection{Vacuum stability, unitarity and perturbativity constraints}

Vacuum stability (boundedness from below) requires the scalar potential to remain positive along every direction in field space at large field values. If all quartic couplings are positive, boundedness holds trivially. However, if at least one mixed quartic coupling (any quartic coupling other than $\lambda_{H,\Phi,S,\eta}$) is negative, additional constraints are needed so the potential does not turn negative at large fields. Following~\cite{Ahriche:2016mcx, Khojali:2022squ}, these conditions take the form
\begin{equation}
\lambda_{H,S,\Phi,\eta}>0,\qquad \det|M_0|>0,\label{eq:VStab}
\end{equation}
where the matrix $M_0$ is defined in Appendix~\ref{app:Unitarity}.

At energies well above the scalar masses, perturbative unitarity 
constrains on the quartic couplings of the scalar potential. using  the
Goldstone boson equivalence theorem, the analysis reduces to scalar
scattering processes $S_{1}S_{2}\to S_{3}S_{4}$, where $S_{i}$ denotes
any scalar degree of freedom defined in Eq.~\eqref{eq:U}. The corresponding
amplitudes are dominated by quartic interactions, and tree-level partial
wave unitarity requires the eigenvalues $\Lambda_{i}$ of the scattering
matrix, given in Appendix~\ref{app:Unitarity}, to satisfy 
\begin{equation}
|\Lambda_{i}|<8\pi.\label{eq:unitarity_bound}
\end{equation}
To ensure perturbativity, we impose the conditions
\begin{align}
&\lambda_H,\lambda_S,\lambda_\Phi,
|\omega_{1,2,4}|,|\lambda_3|,
|\lambda_3+\lambda_4|,|\lambda_\pm|,\nonumber\\
&|\lambda_\eta|,|\omega_3|,|\omega_5|
\leq 4\pi.
\label{eq:perturbativity_bound}
\end{align}

\subsubsection{Electroweak precision tests}

In our setup, the additional gauge bosons $Z^{\prime}$ and $Z^{\prime\prime}$ do not
mix with the SM $Z$ boson at tree level, as shown in Eq.~\eqref{eq:eigen}.
We therefore focus on the one loop contributions of the extended scalar
sector to the oblique parameters $\Delta T$, $\Delta S$, and $\Delta U$,
which are given by 
\begin{widetext}
\begin{equation}
\begin{array}{cc}
\varDelta T= & \frac{1}{16\pi s^{2}_{W}m^{2}_{W}}\left\{ F\Big(m^{2}_{H^{\pm}},m^{2}_{H_{0}}\Big)+F\Big(m^{2}_{H^{\pm}},m^{2}_{A_{0}}\Big)-3\left\{ F\Big(m^{2}_{Z},m^{2}_{H_{1}}\Big)-F\Big(m^{2}_{W},m^{2}_{H_{1}}\Big)\right\} \right.\\
 & -F\Big(m^{2}_{H_{0}},m^{2}_{A_{0}}\Big)+3R^{2}_{11}\left\{ F\Big(m^{2}_{Z},m^{2}_{H_{1}}\Big)-F\Big(m^{2}_{W},m^{2}_{H_{1}}\Big)\right\} \\
 & \left.+3R^{2}_{21}\left\{ F\Big(m^{2}_{Z},m^{2}_{H_{2}}\Big)-F\Big(m^{2}_{W},m^{2}_{H_{2}}\Big)\right\} +3R^{2}_{31}\left\{ F\Big(m^{2}_{Z},m^{2}_{H_{3}}\Big)-F\Big(m^{2}_{W},m^{2}_{H_{3}}\Big)\right\} \right\} ,\\
\varDelta S= & \frac{1}{24\pi}\left\{ \Big(2s^{2}_{W}-1\Big)^{2}G\Big(m^{2}_{H^{\pm}},m^{2}_{H^{\pm}},m^{2}_{Z}\Big)+G\Big(m^{2}_{H_{0}},m^{2}_{A_{0}},m^{2}_{Z}\Big)+\log\Big(\frac{m^{2}_{A_{0}}m^{2}_{H_{0}}}{m^{4}_{H^{\pm}}}\Big)+\right.\\
 & R^{2}_{21}\hat{G}(m^{2}_{H_{2}},m^{2}_{Z})+R^{2}_{31}\hat{G}(m^{2}_{H_{3}},m^{2}_{Z})+R^{2}_{21}\log\Big(m^{2}_{H_{2}}\Big)+R^{2}_{31}\log\Big(m^{2}_{H_{3}}\Big)+\\
 & \left.\Big(R^{2}_{11}-1\Big)\hat{G}(m^{2}_{H_{1}},m^{2}_{Z})+\Big(R^{2}_{11}-1\Big)\log\Big(m^{2}_{H_{1}}\Big)\right\} ,\\
\varDelta U & =\frac{1}{24\pi}\left\{ G\Big(m^{2}_{H^{\pm}},m^{2}_{H_{0}},m^{2}_{W}\Big)+G\Big(m^{2}_{H^{\pm}},m^{2}_{A_{0}},m^{2}_{W}\Big)-G\Big(m^{2}_{H_{0}},m^{2}_{A_{0}},m^{2}_{Z}\Big)-\right.\\
 & \Big(2s^{2}_{W}-1\Big)^{2}G\Big(m^{2}_{H^{\pm}},m^{2}_{H^{\pm}},m^{2}_{Z}\Big)+\Big(R^{2}_{11}-1\Big)\left[\hat{G}\Big(m^{2}_{H_{1}},m^{2}_{W}\Big)-\hat{G}\Big(m^{2}_{H_{1}},m^{2}_{Z}\Big)\right]+\\
 & \left.R^{2}_{21}\left[\hat{G}\Big(m^{2}_{H_{2}},m^{2}_{W}\Big)-\hat{G}\Big(m^{2}_{H_{2}},m^{2}_{Z}\Big)\right]+R^{2}_{31}\left[\hat{G}\Big(m^{2}_{H_{3}},m^{2}_{W}\Big)-\hat{G}\Big(m^{2}_{H_{3}},m^{2}_{Z}\Big)\right]\right\} ,
\end{array}\label{eq:Oblique}
\end{equation}
\end{widetext}

where $s_{W}\equiv\sin\theta_{W}$ is the sine of the weak mixing
angle, and $F$, $G$, and $\hat{G}$ are loop functions defined in
Ref.~\cite{Grimus:2008nb}.

To constrain the model using electroweak precision data, we define
a $\chi^{2}$-function at $1-\sigma$, 
\begin{equation}
\chi^{2}_{EWPT}=\sum_{\mathcal{O}=S,T,U}\frac{(\Delta\mathcal{O}-\Delta\mathcal{O}^{exp})^{2}}{\sigma^{2}_{\mathcal{O}}}<3.53,
\end{equation}
where $\Delta\mathcal{O}$ denotes the theoretical prediction, while
$\Delta\mathcal{O}^{exp}$ and $\sigma_{\mathcal{O}}$ are the experimental
central value and uncertainty, respectively. We adopt $\varDelta S=-0.04\pm0.1,\;\varDelta T=0.01\pm0.12,\:\varDelta U=-0.01\pm0.09$
from Ref.~\cite{ParticleDataGroup:2024cfk}.

\subsubsection{Higgs decay width}

The SM-like Higgs boson $H_{1}$ can decay into pairs of lighter scalars
$S,S'\in\{H_{2},H_{3},H_{0},A_{0}\}$, provided that $m_{S}+m_{S'}<m_{H_{1}}$.
The corresponding partial width is 
\begin{multline}
\Gamma(H_{1}\to SS')=\\
\frac{\lambda^{2}_{H_{1}SS'}\Theta(m_{H_{1}}-m_{S}-m_{S'})}{16\pi(1+\delta_{SS'})m_{H_{1}}}\\
\times\left[1-\frac{2(m^{2}_{S}+m^{2}_{S'})}{m^{2}_{H_{1}}}+\frac{(m^{2}_{S}-m^{2}_{S'})^{2}}{m^{4}_{H_{1}}}\right]^{1/2}.\label{eq:und}
\end{multline}
where $\delta_{SS'}$ accounts for identical final-state particles,
and the trilinear couplings $\lambda_{H_{1}SS'}$ are given in Appendix~\ref{app:SV}. Depending on their decay modes and lifetimes, the new scalars may produce visible or invisible signatures. We denote the width of non-invisible exotic scalar channels by $\Gamma_{\mathrm{und}}=\sum_{(S,S')\in\mathcal{C}_{\mathrm{und}}}\Gamma(H_{1}\to SS')$, where $\mathcal{C}_{\mathrm{und}}$ contains the corresponding scalar pairs, each counted once. For the mass hierarchy considered in our
analysis (see Sect.~\ref{Sec:Numeric}) $m_{H_{2}},m_{H_{3}}>m_{H_{1}}$,
so decays of $H_{1}$ into final states containing $H_{2}$ or $H_{3}$
are kinematically forbidden. 
Including invisible decays, the total Higgs width is 
\begin{equation}
\begin{aligned}\Gamma^{\mathrm{tot}}_{H_{1}}={} & \Gamma_{\mathrm{und}}+\Gamma_{\mathrm{inv}}\\
 & +\Gamma^{\mathrm{SM}}_{H_{1}}\sum_{X=\mathrm{SM}}\kappa^{2}_{X}\,\mathcal{B}^{\mathrm{SM}}(H_{1}\to XX).
\end{aligned}
\label{decayT}
\end{equation}
We impose
$\mathcal{B}_{\mathrm{und}}=\Gamma_{\mathrm{und}}/\Gamma^{\mathrm{tot}}_{H_{1}}<0.22$~\cite{ATLAS:2019nkf,ATLAS:2019cid} and use  the Higgs-width constraint measurement $\Gamma_{H_{1}}=3.7^{+1.9}_{-1.4}~\mathrm{MeV}$~\cite{ParticleDataGroup:2024cfk}. Here $\ensuremath{\Gamma}^{SM}_{H_{1}}=4.02~\mathrm{MeV}$~\cite{Heinemeyer:2013tqa}, and $\kappa_{X}$ denote the corresponding SM-normalized Higgs coupling
modifier, with $\kappa_{V}=\kappa_{F}=R_{11}$ for tree-level couplings
to SM gauge bosons and fermions. 

The invisible Higgs width receives contributions from the kinematically
allowed channels $H_{1}\to SS$, $H_{1}\to VV'$, and $H_{1}\to N_{i}N_{i}$,
provided their final states escape detection. The scalar partial widths
are given in Eq.~\eqref{eq:und}, while those involving vector bosons
and neutral fermions are 
\begin{multline}
\Gamma(H_{1}\to VV')=\frac{|\lambda_{H_{1}VV'}|^{2}}{16\pi(1+\delta_{VV'})m_{H_{1}}}\\
\times\left[1-\frac{2(m^{2}_{V}+m^{2}_{V'})}{m^{2}_{H_{1}}}+\frac{(m^{2}_{V}-m^{2}_{V'})^{2}}{m^{4}_{H_{1}}}\right]^{1/2}\\
\times\left[2+\frac{(m^{2}_{H_{1}}-m^{2}_{V}-m^{2}_{V'})^{2}}{4m^{2}_{V}m^{2}_{V'}}\right],
\end{multline}
and 
\begin{equation}
\Gamma(H_{1}\to N_{i}N_{i})=\frac{|\lambda_{H_{1}N_{i}N_{i}}|^{2}m_{H_{1}}}{16\pi}\left(1-\frac{4M^{2}_{i}}{m^{2}_{H_{1}}}\right)^{3/2},
\end{equation}
with $\lambda_{H_{1}VV'}g^{\mu\nu}$ denoting the $H_{1}V_{\mu}V'_{\nu}$
vertex. The total invisible width is 
\begin{equation}
\begin{aligned}\Gamma_{\mathrm{inv}}={} & \sum^{\mathrm{inv}}_{S=H_{0},A_{0}}\Gamma(H_{1}\to SS)+\sum^{\mathrm{inv}}_{V,V'}\Gamma(H_{1}\to VV')\\
 & +\sum^{\mathrm{inv}}_{i}\Gamma(H_{1}\to N_{i}N_{i}),
\end{aligned}
\end{equation}
where the sums include allowed channels, involving stable particles
or particles decaying outside the detector, and the invisible branching
ratio must satisfy the ATLAS bound $\mathcal{B}_{\mathrm{inv}}=\Gamma_{\mathrm{inv}}/\Gamma^{\mathrm{tot}}_{H_{1}}\leq0.11$~\cite{ATLAS:2020kdi}.

\subsubsection{The diphoton Higgs decay}

We evaluate the signal strength for $h\to\gamma\gamma$ as 
\begin{equation}
\begin{gathered}\mu_{\gamma\gamma}=\frac{\sigma(pp\to H_{1})\mathcal{B}(H_{1}\to\gamma\gamma)}{\sigma^{\mathrm{SM}}(pp\to H_{1})\mathcal{B}^{\mathrm{SM}}(H_{1}\to\gamma\gamma)}\\
=\kappa^{2}_{F}\frac{\Gamma^{\mathrm{SM}}_{H_{1}}}{\Gamma^{\mathrm{tot}}_{H_{1}}}\left|\frac{\mathcal{A}_{\gamma\gamma}}{\mathcal{A}^{\mathrm{SM}}_{\gamma\gamma}}\right|^{2},\\
\mathcal{A}_{\gamma\gamma}=\frac{\lambda_{H_{1}H^{+}H^{-}}\upsilon}{2m^{2}_{H^{\pm}}}A^{\gamma\gamma}_{0}(\tau_{H^{\pm}})\\
\qquad+\kappa_{V}A^{\gamma\gamma}_{1}(\tau_{W})+\kappa_{F}\sum_{f}N^{f}Q^{2}_{e}A^{\gamma\gamma}_{1/2}(\tau_{f}),\\
\mathcal{A}^{\mathrm{SM}}_{\gamma\gamma}=A^{\gamma\gamma}_{1}(\tau_{W})+\sum_{f}N^{f}Q^{2}_{e}A^{\gamma\gamma}_{1/2}(\tau_{f}).
\end{gathered}
\end{equation}
Here, $\sigma(pp\to H_{1})$ and $\sigma(pp\to H_{1})^{SM}$ denote
the gluon-gluon fusion (ggF) production cross section in the model
and the SM, respectively, while $\mathcal{B}$ and $\mathcal{B}^{SM}$
represent the corresponding branching ratios; $\lambda_{H_{1}H^{+}H^{-}}$
is tree level scalar trilinear coupling defined in Appendix~\ref{app:SV},
and $N^{f}$ and $Q^{f}_{e}$ are the color factor and electric charge
of the fermions, respectively. The functions $A^{\gamma\gamma}_{0,1/2,1}$
are given in the literature~\cite{Chen:2013vi}.

\subsubsection{Higgs Signal strengths}

We consider Higgs production through ggF, for which the signal strength
in the decay channel $H_{1}\to XX$ is given in the narrow-width approximation
(NWA) by 
\begin{equation}
\begin{gathered}\mu_{XX}=\frac{\sigma(pp\to H_{1}\to XX)}{\sigma^{\mathrm{SM}}(pp\to H_{1}\to XX)}\\
=\frac{\sigma(pp\to H_{1})}{\sigma^{\mathrm{SM}}(pp\to H_{1})}\frac{B(H_{1}\to XX)}{B^{\mathrm{SM}}(H_{1}\to XX)}\\
=\kappa^{2}_{F}\kappa^{2}_{X}\frac{\Gamma^{\mathrm{SM}}_{H_{1}}}{\Gamma^{\mathrm{tot}}_{H_{1}}}.
\end{gathered}
\end{equation}
where $\sigma(pp\to H_{1})$ and $\sigma^{\mathrm{SM}}(pp\to H_{1})$
denote the ggF cross sections in our model and SM, respectively, while
$\kappa^{2}_{X}$ denotes the SM-normalized partial-width ratio, and
$\Gamma^{\mathrm{tot}}_{H_{1}}$ is given in Eq.~\eqref{decayT}.
For the diphoton channel, $\kappa^{2}_{\gamma}$ includes the charged-scalar
loop contribution discussed above. To assess agreement with the measured
Higgs signal strengths in the $\gamma\gamma$, $b\bar{b}$, $WW$,
$ZZ$, $\tau^{+}\tau^{-}$, and $\mu^{+}\mu^{-}$ channels, we define
\begin{equation}
\chi^{2}_{SM}=\sum_{X}\frac{\left(\mu_{XX}-\mu^{exp}_{XX}\right)^{2}}{\left(\Delta\mu^{exp}_{XX}\right)^{2}},
\end{equation}
Using the experimental central values $\mu^{\mathrm{exp}}_{XX}$ and
uncertainties $\Delta\mu^{\mathrm{exp}}_{XX}$ compiled in Ref.~\cite{ParticleDataGroup:2024cfk},
we retain only benchmark points satisfying the $1-\sigma$ constraint:
$\chi^{2}_{\mathrm{SM}}<7.04$.

\subsubsection{LHC constraints on $Z^{\prime}$ and $Z^{\prime\prime}$}

The interactions of the additional neutral gauge bosons $Z^{\prime}$ and
$Z^{\prime\prime}$ with SM fermions, relevant to LHC dilepton and dijet resonance
searches, are described by 
\begin{equation}
\mathcal{L}=\sum_{V=Z^{\prime},Z^{\prime\prime}}\sum_{f}\bar{f}\gamma^{\mu}\left(a^{V,f}_{v}+a^{V,f}_{a}\gamma^{5}\right)f\,V_{\mu},
\end{equation}
where $a^{V,f}_{v}$ and $a^{V,f}_{a}$ denote the vector and axial-vector
couplings, respectively. For comparison, the corresponding SM $Z$
couplings are $a^{Z,f}_{v}=-g(T^{f}_{3}-2Q_{f}s^{2}_{W})/(2c_{W})$
and $a^{Z,f}_{a}=gT^{f}_{3}/(2c_{W})$, with $c_{W}\equiv\cos\theta_{W}$.
In the narrow-width approximation (NWA), the dilepton resonance constraints
can be expressed in terms of the production cross-section ratio 
\begin{align}
R^{V}_{\sigma} & =\frac{\sigma(pp\xrightarrow{V}f\bar{f})}{\sigma^{\mathrm{SM}}(pp\xrightarrow{Z}f\bar{f})}=\frac{\sigma(pp\to V)}{\sigma^{\mathrm{SM}}(pp\to Z)}\,\frac{B_{m_{V}}(V\to f\bar{f})}{B^{\mathrm{SM}}_{m_{Z}}(Z\to f\bar{f})}\nonumber \\
 & =\frac{\kappa^{2}_{u}\sigma_{u}(pp\to V)+\kappa^{2}_{d}\sigma_{d}(pp\to V)}{\sigma^{\mathrm{SM}}_{m_{Z}}(pp\to Z)}\,\frac{B_{m_{V}}(V\to f\bar{f})}{B^{\mathrm{SM}}_{m_{Z}}(Z\to f\bar{f})},\label{eq:RV}
\end{align}
The quantities $\sigma_{u,d}(m_{V})$ denote the reference production cross
sections from up and down type quark annihilation, evaluated at $m_{V}$
with SM $Z$ couplings. The coupling modifiers are defined as 
\begin{equation}
\kappa^{2}_{V,f}=\frac{(a^{V,f}_{v})^{2}+(a^{V,f}_{a})^{2}}{(a^{Z,f}_{v})^{2}+(a^{Z,f}_{a})^{2}}\label{eq:kappa_V}
\end{equation}
with further details given in Appendix~\ref{app:kaapa}. The branching
ratio is $\mathcal{B}(V\to f\bar{f})=\Gamma(V\to f\bar{f})/\Gamma^{\mathrm{tot}}_{V}$,
where the total width includes all kinematically allowed SM and beyond
standard model (BSM) channels, 
\begin{equation}
\Gamma^{\mathrm{tot}}_{V}=\sum_{f\in\mathrm{SM}}\Gamma(V\to f\bar{f})+\Gamma(V\to W^{+}W^{-})+\Gamma_{V\to\mathrm{BSM}},\label{eq:GV}
\end{equation}
The fermionic partial width is 
\begin{multline}
\Gamma(V\to f\bar{f})=\frac{N_{f}m_{V}}{12\pi}\sqrt{1-\frac{4m^{2}_{f}}{m^{2}_{V}}}\\
\times\left[(a^{V,f}_{v})^{2}\left(1+\frac{2m^{2}_{f}}{m^{2}_{V}}\right)+(a^{V,f}_{a})^{2}\left(1-\frac{4m^{2}_{f}}{m^{2}_{V}}\right)\right],\label{eq:Vff}
\end{multline}
where $N_{f}=3$ for quarks and $N_{f}=1$ for leptons. For a Yang-Mills
triple-gauge vertex, defining $x=m^{2}_{W}/m^{2}_{V}$, the $W^{+}W^{-}$
partial width reads 
\begin{equation}
\begin{aligned}\Gamma(V\to W^{+}W^{-})={} & \frac{|g_{VWW}|^{2}m_{V}}{192\pi}\frac{(1-4x)^{3/2}}{x^{2}}\\
 & \times(1+20x+12x^{2}),
\end{aligned}
\label{eq:VWW}
\end{equation}
In our setup, the absence of tree level mixing between $Z$ and the
additional gauge bosons $Z^{\prime}$ and $Z^{\prime\prime}$ implies $g_{VWW}=0$ for
$V=Z^{\prime},Z^{\prime\prime}$, so the $W^{+}W^{-}$ channel does not contribute to their
total decay widths at tree level. For decays into pair of Majorana
fermions $N_{i}$, the corresponding partial width is 
\begin{equation}
\Gamma(V\to N_{i}N_{i}) = \frac{g^{2}_{VN_{i}}m_{V}}{24\pi}\left(1-\frac{4m^{2}_{N_{i}}}{m^{2}_{V}}\right)^{3/2}.
\end{equation}
where $g_{VN_{i}}$ is defined through the axial interaction $-g_{VN_{i}}V_{\mu}\bar{N}_{i}\gamma^{\mu}\gamma^{5}N_{i}/2$.
The decay into a lighter neutral gauge boson and a CP-even scalar
is given by
\begin{multline}
\Gamma(V'\to VH_{i})=\frac{\lambda^{2}_{H_{i}VV'}\lambda^{1/2}(m^{2}_{V'},m^{2}_{V},m^{2}_{H_{i}})}{48\pi m^{3}_{V'}}\\
\times\left[2+\frac{(m^{2}_{V'}+m^{2}_{V}-m^{2}_{H_{i}})^{2}}{4m^{2}_{V'}m^{2}_{V}}\right].
\end{multline}
Here $\lambda_{H_{i}VV'}g^{\mu\nu}$ denoting the $H_{i}V_{\mu}V'_{\nu}$
vertex, and $\lambda(x,y,z)=x^{2}+y^{2}+z^{2}-2xy-2xz-2yz$ is the
Källén function. Additional BSM decays, such as $V\to H_{0}A_{0}$,
would increase $\Gamma^{\mathrm{tot}}_{V}$, suppressing the dilepton and
dijet branching fractions and modifying the LHC constraints, however, this is not the case as the vertices $Z^{\prime}\to H_{0}A_{0}$ and $Z^{\prime\prime}\to H_{0}A_{0}$ are strictly vanishing.

\section{Dark Matter Candidates~\label{Sec:DM}}

The model admits several DM candidates, which must be electrically
neutral, cosmologically stable, and consistent with the observed relic
density and direct-detection constraints. The exact $Z_{2}$ symmetry
stabilizes the lightest $Z_{2}$-odd particle among the components
of the inert doublet $\eta$ and the Majorana fermions $N_{i}$, while
all SM fields are $Z_{2}$ even.

\subsection{Fermionic and inert-doublet DM}

If $m_{N_{1}}<m_{H_{0}},\,m_{A_{0}},\,m_{H^{\pm}}$, the lightest
singlet Majorana fermion, $N_{1}$, is stable and can serve as a fermionic
DM candidate, as in the scotogenic model~\cite{Ahriche:2017iar}.
Its annihilation channels mediated by the scotogenic Yukawa interactions
include 
\begin{equation}
N_{1}N_{1}\to\ell^{+}_{\alpha}\ell^{-}_{\beta},\qquad N_{1}N_{1}\to\nu_{\alpha}\nu_{\beta},
\end{equation}
which proceed through the $t$-channel exchange of $H^{\pm}$, $H_{0}$,
and $A_{0}$ through the Yukawa interaction $f_{i\alpha}\overline{L}_{\alpha}\widetilde{\eta}N_{i}$.
Additional channels can proceed through $s$-channel exchange of the
scalar bosons $H_{i}$ and the neutral gauge bosons $Z^{\prime}$ and $Z^{\prime\prime}$:
\begin{align}
N_{1}N_{1} &\to H^{*}_{i}\to
W^{+}W^{-},\,ZZ,\,H_{j}H_{k},\,
Q^{0}\bar{Q}^{0},\,f\bar{f},\\
N_{1}N_{1} &\to Z^{\prime *},Z^{\prime\prime *}
\to Q^{0}\bar{Q}^{0},\,f\bar{f}.
\end{align}
whenever kinematically allowed, with $H_{i}=(H_{1},H_{2},H_{3})$.
These additional gauge-boson-mediated channels distinguish the present
framework from the scale-invariant scotogenic model~\cite{Soualah:2021xbn}.
Resonant annihilation may significantly enhance the effective cross
section near 
\begin{equation}
m_{N_{1}}\simeq\frac{m_{Z^{\prime}}}{2},\qquad\frac{m_{Z^{\prime\prime}}}{2},\qquad\frac{m_{H_{i}}}{2}.
\end{equation}

The neutral inert scalars $H_{0}$ and $A_{0}$ can each serve as
a stable DM candidates, with stability guaranteed for the lightest $Z_{2}$-odd state. Their mass splitting is governed by $\lambda_{5}$, 
\begin{equation}
m^{2}_{H_{0}}-m^{2}_{A_{0}}=\lambda_{5}v^{2}.
\end{equation}
Thus, for $\lambda_{5}<0$ ($\lambda_{5}>0$), $H_{0}$ ($A_{0}$)
is the lighter state and can annihilate through channels such as 
\begin{equation}
\begin{aligned}H_{0}H_{0}\,(A_{0}A_{0})\to & W^{+}W^{-},\,ZZ,\\
H_{0}H_{0}\,(A_{0}A_{0})\to & f\overline{f},
\end{aligned}
\end{equation}
through gauge and Higgs-portal interactions, respectively, while scalar
final states such as $H_{i}H_{j}$ may also contribute when kinematically
accessible.

For sizable Yukawa couplings $f_{\alpha i}$, the radiative-neutrino-mass
constraint generally requires a small $\lambda_{5}$, typically of
order $\lambda_{5}\sim10^{-9}$ in the parameter regions of interest.
Consequently, $H_{0}$ and $A_{0}$ can be nearly degenerate, making
co-annihilation important. Relevant processes include 
\begin{equation}
H_{0}A_{0}\to f\overline{f},\qquad H_{0}H^{\pm}\to W^{\pm}Z,\;W^{\pm}\gamma,
\end{equation}
and analogous channels involving $A_{0}H^{\pm}$. For inert scalar
DM, however, the same neutral scalar mass splitting controls $Z$-mediated
inelastic scattering, $H_{0}\mathcal{N}\to A_{0}\mathcal{N}$,the LZ comparison presented here is restricted to elastic scalar-mediated scattering; the inelastic channel is not included in this analysis.

\subsection{Vector DM}

The complex hidden gauge boson $Q^{0}_{\mu}=\frac{1}{\sqrt{2}}\big(Q^{1}_{\mu}+iQ^{2}_{\mu}\big)$
does not couple directly to SM fermions and can provide a natural
vector-DM candidate, analogous to vector DM in hidden non-Abelian
gauge theories~\cite{Baouche:2021wwa}. Its stability is ensured by a residual discrete symmetry of the hidden gauge sector combined with fermion parity,  defined as
\begin{equation}
{\cal P}_{Q} = (-1)^{3\left(T^{H}_3 + n_1 - n_2\right) +  2 j}
\end{equation}
where $T^{H}_3$ is the diagonal generator of $SU(2)_H$ and $j$ is the spin of the field on which ${\cal P}_{Q}$ acts. Under this symmetry, $Q^{0}$ and $\bar  Q^{0}$  are the only odd physical states and are therefore stable independently of the imposed $Z_2$ symmetry.
Interactions involving a $Q^{0}\overline{Q}^{\,0}$ pair remain allowed, including
\begin{equation}
Q^{0}\overline{Q}^{\,0}H_{i},\qquad Q^{0}\overline{Q}^{\,0}Z^{\prime},\qquad Q^{0}\overline{Q}^{\,0}Z^{\prime\prime}.
\end{equation}
The corresponding annihilation channels include 
\begin{align}
Q^{0}\overline{Q}^{\,0} & \to H^{*}_{i}\to W^{+}W^{-},\,ZZ,\nonumber \\
Q^{0}\overline{Q}^{\,0} & \to H^{*}_{i},Z^{\prime},Z^{\prime\prime}\to f\overline{f},\,N_{i}N_{i},\,H_{j}H_{k},\nonumber \\
Q^{0}\overline{Q}^{\,0} & \to Z^{\prime *},Z^{\prime\prime *}\to f\overline{f},\,N_{i}N_{i},\nonumber \\
Q^{0}\overline{Q}^{\,0} & \to Z^{\prime}Z^{\prime},\,Z^{\prime\prime}Z^{\prime\prime},\,Z^{\prime}Z^{\prime\prime},
\end{align}
when these final states are kinematically open. Resonant enhancement
can occur for 
\begin{equation}
m_{Q^{0}}\simeq\frac{m_{H_{i}}}{2},\qquad\frac{m_{Z^{\prime}}}{2},\qquad\frac{m_{Z^{\prime\prime}}}{2}.\label{eq:mq0:1}
\end{equation}
The neutral gauge bosons $Z^{\prime}$ and $Z^{\prime\prime}$ are not DM candidates because
they couple to fermions and are not protected by the residual symmetry.
In the hierarchical regime $v\ll v_{S},v_{\Phi}$, one obtains 
\begin{equation}
m_{Z^{\prime}}\lesssim m_{Q^{0}}<m_{Z^{\prime\prime}}.
\end{equation}
In this regime, the $Z^{\prime}$ pole in \eqref{eq:mq0:1} can not be reached
in non-relativistic $Q^{0}\overline{Q}^{\,0}$ annihilation. Moreover, since $Q^0$ is parity odd and $Z'$ is parity eve, an initial state containing both particles can not produce only parity even final state. 
For example, a scattering  of the form 
\begin{equation}
Z^{\prime}+Q^{0}\to Q^{0}+X,
\end{equation}
with $X$ containing no additional stable DM particle, does not change
the total number of DM particles. Such reactions may affect the kinetic
or chemical-equilibrium history, but they do not constitute ordinary
co-annihilation that depletes the $Q^{0}$ abundance.

\subsection{Multi-component DM}

If both $Q^{0}$ and another stable state $\chi$ are present, with
$\chi=N_{1}$, $H_{0}$, or $A_{0}$, the relic abundance is 
\begin{equation}
\Omega_{{\rm DM}}h^{2}=\Omega_{Q^{0}}h^{2}+\Omega_{\chi}h^{2}.
\end{equation}
Individual contributions do not need to reproduce the observed abundance
separately; only their sum must satisfy $\Omega_{{\rm DM}}h^{2}\simeq0.12$.
For illustration, requiring $Q^{0}$ to account for at most $95\%$
of the observed abundance implies $\Omega_{Q^{0}}h^{2}\lesssim0.114$,
while the second component contributes at least approximately $0.006$.
These numerical fractions are benchmark choices rather than model-independent
requirements.

The relic density of $Q^{0}$ is controlled by its self-annihilation
and, where relevant, by conversion processes involving the other stable
component. A rough estimate based on the canonical thermal cross section,
$\langle\sigma v\rangle_{{\rm th}}\sim3\times10^{-26}\,{\rm cm^{3}\,s^{-1}}$,
suggests that 
\begin{equation}
\langle\sigma(Q^{0}\overline{Q}^{\,0})v\rangle\gtrsim\frac{\langle\sigma v\rangle_{{\rm th}}}{0.95},
\end{equation}
if $Q^{0}$ supplies no more than $95\%$ of the total relic density.
Because annihilation and conversion processes can couple to the two
stable abundances, this estimate is not a general bound on $g_{H}$;
deriving such a bound requires solving the coupled Boltzmann equations
with the relevant channels included.

\par\vspace{2.5pt}
For multicomponent DM, the predicted direct-detection
rates and indirect-detection annihilation signals are
obtained by summing the contributions from each
component~\cite{Zurek:2008qg,Herrero-Garcia:2017vrl,Profumo:2009tb},
\begin{equation}
\mathcal{O}_{\mathrm{tot}}
=\sum_{\chi}\xi_{\chi}^{n}\mathcal{O}_{\chi}^{(1)},
\quad
n=
\begin{cases}
1, & \text{direct detection},\\
2, & \text{self-annihilation}.
\end{cases}
\end{equation}
Here, $\xi_{\chi}=\Omega_{\chi}/\Omega_{\mathrm{DM}}^{\mathrm{tot}}$
is the fractional DM abundance, and
$\mathcal{O}_{\chi}^{(1)}$ denotes the predicted
signal assuming $\xi_{\chi}=1$.

\section{DM Direct Detection~\label{sec:DM_DD}}

\begin{figure}[ht]
\begin{centering}
\includegraphics[width=0.95\columnwidth]{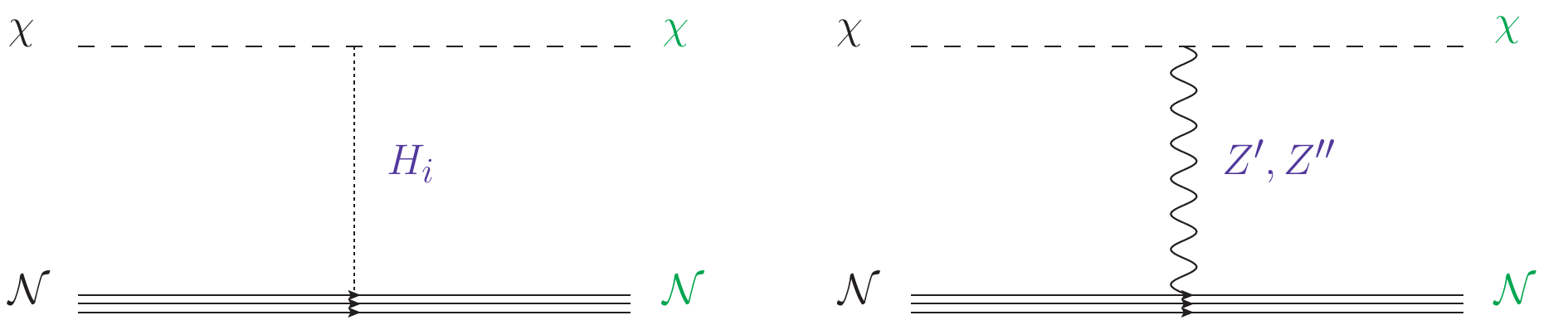} 
\par\end{centering}
\caption{Feynman diagrams that are responsible for DM direct detection. Here, DM candidate $\chi$ represents $N_{1}$, $H_{0}$ and $Q^{0}$; and
the left diagram applies only for $N_{1}$ and $Q^{0}$.}
\label{fig:DD} 
\end{figure}

We now examine the scattering of the DM candidates off nucleons through
scalar and gauge boson exchange and assess the corresponding constraints
from direct detection experiments. The tree level contributions are
illustrated in Fig.~\ref{fig:DD}, with CP-even scalars $H_{i}~(i=1,3)$
mediating the elastic scattering of all three candidates, while $Z^{\prime}$
and $Z^{\prime\prime}$ exchange contribute to the $N_{1}$ and $Q^{0}$.  

By combining the relevant DM interaction terms involving
($H_{k},~Z^{\prime},~Z^{\prime\prime}$) and integrating out these mediators at the
low momentum transfers characteristic of direct detection, one obtains
the effective Lagrangian 
\begin{widetext}
\begin{align}
\mathcal{L}_{{\rm eff}} & =-\frac{g_{H}m_{Q^{0}}m_{q}S_{13}}{\upsilon}\,\bar{Q}^{0}_{\mu}Q^{0\mu}\bar{q}q-\frac{g_{H}g_{1}s_{\alpha}c_{\alpha}}{3}\Big(\frac{1}{m^{2}_{Z^{\prime}}}-\frac{1}{m^{2}_{Z^{\prime\prime}}}\Big)\,(p+p')^{\mu}\left(-\epsilon'^{*}\cdot\epsilon\right)\,\bar{q}\gamma_{\mu}q\nonumber \\
 & -\frac{m_{q}\lambda_{H_{k}H_{0}H_{0}}R_{k1}}{2\upsilon\,m^{2}_{H_{k}}}H_{0}H_{0}\bar{q}q-\frac{m_{N_{i}}m_{q}S_{12}}{2\upsilon\upsilon_{S}}\,\bar{N}_{i}N_{i}\bar{q}q+\frac{g^{2}_{1}}{6}\Big(\frac{s^{2}_{\alpha}}{m^{2}_{Z^{\prime}}}+\frac{c^{2}_{\alpha}}{m^{2}_{Z^{\prime\prime}}}\Big)\bar{N}_{i}\gamma^{\mu}\gamma_{5}N_{i}\,\bar{q}\gamma_{\mu}q,\label{eq:effL}
\end{align}
with $S_{ij}\equiv\sum_{k}\frac{R_{ki}R_{kj}}{m^{2}_{H_{k}}}$, with
summation over quark flavors and repeated scalar indices understood.
\end{widetext}

To estimate the direct detection cross section, one needs the nucleon
matrix elements. For light quarks, we use $\sum_{u,d,s}m_{q}\langle\bar{q}q\rangle=m_{\mathcal{N}}-m_{\mathcal{B}}$,
with $m_{\mathcal{N}}$ is the nucleon mass and $m_{\mathcal{B}}$
denotes its value in the three flavor chiral limit~\cite{Hoferichter:2017olk}.
For heavy quarks, we have $\sum_{c,b,t}m_{q}\langle\bar{q}q\rangle=\tfrac{2}{9}m_{\mathcal{B}}$.
The resulting effective Higgs--nucleon scalar coupling is therefore~\cite{He:2008qm}
\begin{equation}
F_{N}\equiv f_{N}\; m_{\mathcal{N}},\quad f_{N} = 1 - \frac{7}{9} \frac{m_{\mathcal{B}}}{m_{\mathcal{N}}}\approx0.3.
\end{equation}
We define the  reduced mass of particles $A$ and $B$ as $\mu_{AB}\equiv m_A m_B/(m_A+m_B)$. For scalar DM, the process $H_{0}\mathcal{N}\to H_{0}\mathcal{N}$ occurs only through the exchange of  $H_{k}$ only since $H_{0}$ has no diagonal coupling to the  $Z^{\prime}/Z^{\prime\prime}$ gauge bosons; its cross section in the non-relativistic (NR) limit is given by\footnote{For $A_{0}$ as DM, the elastic cross section follows by replacing $H_{0}$ by $A_{0}$ and using the appropriate trilinear couplings, while the off-diagonal $ZH_{0}A_{0}$ interaction allows inelastic scattering if kinematically allowed by the $H_{0}-A_{0}$ mass splitting.}

\begin{equation}
\sigma_{H_{0}\mathcal{N}}= \frac{\mu^2_{H^0 {\cal N}}}{4\pi}\frac{F^{2}_{N}}{\,m^{2}_{H_{0}}\,\upsilon^{2}}\left[\sum_{k}\frac{\lambda_{H_{k}H_{0}H_{0}}R_{k1}}{m^{2}_{H_{k}}}\right]^{2}.
\end{equation}

For the scattering process $N_{1}\mathcal{N}\to N_{1}\mathcal{N}$, the  contribution from $\gamma^{5}$ term (last term in Eq.~(\ref{eq:effL})) is velocity-suppressed in
the NR limit, with $v\sim10^{-3}$, and can be neglected when evaluating
the leading spin-independent scalar contribution, yielding the cross
section 
\begin{equation}
\sigma_{N_{i}\mathcal{N}}= \frac{\mu^2_{N_1 {\cal N}}}{\pi}\frac{F^{2}_{N} m^{2}_{N_{1}}}{\upsilon^{2}\upsilon^{2}_{s}}\,S^{2}_{12}.
\end{equation}

The complex vector DM, the elastic scattering process $Q^{0}\mathcal{N}\to Q^{0}\mathcal{N}$ and $\bar Q^{0}\mathcal{N}\to \bar Q^{0}\mathcal{N}$
receive contributions from scalar ($H_{k}$) and heavy gauge boson ($Z^{\prime}/Z^{\prime\prime}$) exchange. The scalar contribution is the same for $Q^0$ and $\bar Q^0$. The corresponding cross sections are 
\begin{align}
\sigma_{Q^{0}\mathcal{N}} &= \frac{g^{2}_{H}\mu^2_{Q^0 {\cal N}}}{4\pi} \left[ \frac{F_{\mathcal{N}}}{\upsilon}S_{13} -2g_{1}s_{\alpha}c_{\alpha} \Big(\frac{1}{m^{2}_{Z^{\prime}}} -\frac{1}{m^{2}_{Z^{\prime\prime}}}\Big) \right]^{2},\\ 
\sigma_{\bar Q^{0}\mathcal{N}} &=\frac{g^{2}_{H}\mu^2_{Q^0 {\cal N}}}{4\pi}
\left[ \frac{F_{\mathcal{N}}}{\upsilon}S_{13} +2g_{1}s_{\alpha}c_{\alpha}
\Big(\frac{1}{m^{2}_{Z^{\prime}}} -\frac{1}{m^{2}_{Z^{\prime\prime}}}\Big)
\right]^{2}.
\end{align}

For equal local densities of $Q^0$ and $\bar Q^0$, i.e. assuming no  particle-antiparticle asymmetry,  the effective cross section is simply the average of the two contributions, i.e.  
\begin{align}
\sigma^{(\text{eff})}_{Q\mathcal{N}} &= \frac{\sigma_{Q^{0}\mathcal{N}}
+ \sigma_{\bar Q^{0}\mathcal{N}}}{2} \nonumber\\
&=\frac{g^{2}_{H}\mu^2_{Q^0\mathcal{N}}}{4\pi}
\left[
\frac{F^2_{\mathcal{N}}}{\upsilon^2}S^2_{13}
+4g^2_{1}s^2_{\alpha}c^2_{\alpha}
\Big(\frac{1}{m^{2}_{Z^{\prime}}}
-\frac{1}{m^{2}_{Z^{\prime\prime}}}\Big)^2
\right].
\label{eq:sigmaQeff}
\end{align}

The cross sections $\sigma_{N_1\mathcal{N}}$,  $\sigma_{H^0\mathcal{N}}$, and $\sigma^{(\text{eff})}_{Q {\cal N}}$ are compared separately with the  LZ (2025) experimental limit~\cite{LZ:2024zvo} . Since the relic density analysis is deferred to a future work~\cite{next}, each comparison assumes that the candidate accounts for the entire  local DM density ($\xi_{\chi} = 1$) with $Q^0$ and $\bar Q^0$ counted together. A combined constraint requires the component fractions and the sum of their predicted  nuclear recoil rates. 

The LZ (2025) experiment reported an event consistent with  nuclear-recoil candidate at about $248~\mathrm{keV}$, with a global significance of $2.6\sigma$~\cite{Arhrib:2013ela}. In our model, inelastic scattering mediated by the $Z$ boson, in which the lighter inert scalar ($H^0$ or $A^0$) converts into the heavier one, could produce high-energy nuclear recoils. However, determining whether this process can account for the reported event requires a detailed analysis of the recoil spectrum and event rate, taking into account the detector response, the local scalar DM abundance, and existing experimental constraints, which is beyond the scope of the present work.

\section{DM Indirect Detection~\label{sec:DM_ID}}

Indirect detection experiments search for gamma rays, cosmic rays
(antiprotons, positrons), and neutrinos produced by DM annihilation
in environments such as the Galactic Center and dwarf spheroidal galaxies~\cite{Slatyer:2017sev,Gaskins:2016cha}.
The nature and observability of indirect detection signals depend
on the dominant annihilation channels and the couplings of these candidates
to the SM.

A continuum signal arises from annihilation into SM particles, such
as $\chi\chi\to f\bar{f},\quad W^{+}W^{-}$, and $ZZ$, or into scalar
pairs $H_{i}H_{j}$ that subsequently decay into visible SM states.
These processes can produce continuum gamma rays and charged cosmic
rays (positrons and antiprotons) through hadronization and decay of
the primary final states, with spectra that can be computed using
standard tools such as PPPC4DMID~\cite{Cirelli:2010xx}. The Majorana
particle $N_{1}$ annihilates through Yukawa interactions into charged
leptons or neutrinos; inert scalar annihilation into $W^{+}W^{-}$
and $ZZ$ analogously to the inert doublet model~\cite{LopezHonorez:2006gr},
while $Q^{0}\overline{Q}^{\,0}$ pairs can annihilate through scalar
exchange and, for fermionic final states, through $Z^{\prime}$ and $Z^{\prime\prime}$ exchange.

Monochromatic gamma-ray lines, arising from loop-induced processes
such as $\chi\chi\to\gamma\gamma$ and $\chi\chi\to\gamma Z$, provide
a distinctive signature of DM annihilation~\cite{Ibarra:2012dw}.
For inert scalar DM, $H_{0}H_{0}\to\gamma\gamma/\gamma Z$ can proceed
at one loop through $W^{\pm}$ and $H^{\pm}$ exchange~\cite{Gustafsson:2007pc,Arhrib:2013ela},
giving $E_{\gamma}\simeq m_{H_{0}}$ for the $\gamma\gamma$ channel
and $E_{\gamma}\simeq m_{H_{0}}\left(1-m^{2}_{Z}/4m^{2}_{H_{0}}\right)$
for the $\gamma Z$ channel. For Majorana $N_{1}$, analogous lines
arise from charged scalar and charged lepton loops~\cite{Bergstrom:1989jr,Bringmann:2007nk};
although LFV constrain the underlying Yukawa couplings, they do not
by themselves imply that these loop induced signal are negligible.

Annihilation into states such as $Z^{\prime}Z^{\prime}$, $Z^{\prime\prime}Z^{\prime\prime}$, $H_{0}H_{0}$,or
$A_{0}A_{0}$, can produce cascade signals when kinematically accessible.
If these states subsequently decay into SM particles, the resulting
gamma-ray is broad and  forms a continuum, potentially
with box-shaped features for suitable decay kinematics, whereas stable
or sufficiently long-lived states suppress the decay-induced signal.

The annihilation cross-section relevant for indirect detection depends
on the velocity distribution of DM particles. In general, s-wave annihilation
yields a leading velocity-independent contribution, while p-wave annihilation
is suppressed by $v^{2}$. For $N_{1}N_{1}\to \ell^{+}\ell^{-}$ through
inert scalar exchange, the two body s-wave contribution is helicity-suppressed
for light charged leptons. Astrophysical environments probe different
velocity regimes, with typical velocities $v\sim10^{-5}-10^{-4}$
in dwarf spheroidal galaxies and $v\sim10^{-3}$ in the Milky Way
halo, making the relative importance of s-wave and p-wave contributions
channel and target dependent.

Observations of gamma rays from dwarf spheroidal galaxies by the \textit{Fermi}-LAT
collaboration provide some of the most stringent constraints on DM
annihilation~\cite{Fermi-LAT:2015att}. For annihilation into $b\bar{b}$
or $W^{+}W^{-}$, the bounds can approach the canonical thermal relic
cross section, $\langle\sigma v\rangle\sim3\times10^{-26}\,\mathrm{cm}^{3}\mathrm{s}^{-1}$,
for DM masses below a few hundred GeV. Complementary constraints arise
from measurements of cosmic-ray antiprotons by AMS-02~\cite{AMS:2016oqu,Cirelli:2013hv},
which are particularly sensitive to hadronic final states, although
they depend on cosmic-ray propagation assumptions. In addition , the
energy injection from DM annihilation during recombination is constrained
by observations of the cosmic microwave background (CMB) by the \textit{Planck}
satellite.

For complex vector DM,  the annihilation rate depends on the product of the $Q^0$ and $\bar Q^0$ densities, while CMB constraints require their respective recombination-era abundances and the appropriate normalization.
 Since quantitative indirect-detection limits require channel-specific annihilation rates and appropriate abundance weighting, a numerical assessment of these constraints across the viable parameter space is beyond the scope of the present analysis and will be pursued in a future work~\cite{next}.

\section{Numerical Results~\label{Sec:Numeric}}

In our numerical analysis, we perform a random parameter scan over
the set of model input parameters: 
\begin{itemize}
\item Scalar masses: $m_{H_{2}}$, $m_{H_{3}}$, $m_{H_{0}}$, $m_{A_{0}}$,
and $m_{H^{\pm}}$; 
\item Gauge boson masses and couplings: $m_{Z^{\prime}}$ and dark gauge
coupling $g_{H}$ ; 
\item Majorana neutrino masses: $m_{N_{1}}$, $m_{N_{2}}$, and $m_{N_{3}}$
; 
\item Scalar mixing angles $s_{12}$, $s_{13}$, and $s_{23}$, along with
neutral gauge mixing parameter $s_{\alpha}$ ; 
\item Scalar potential parameters: $\omega_{3}$, $\omega_{5}$, $\mu^{2}_{\eta}$, and $\lambda_{\eta}$. 
\end{itemize}
All points are subject to vacuum stability, perturbativity, and unitarity
requirements (see Sec.~\ref{Sec:Constraints}). The VEVs $v_{s}$
and $v_{\Phi}$, as well as the heavy neutral gauge boson mass $m_{Z^{\prime\prime}}$,
are dynamically determined from $m_{Z^{\prime}}$, $g_{H}$, and $s_{\alpha}$
using the relations in Sec.~\ref{Sec:Model}. Additionally, we restrict
the VEV ranges to $10~\text{GeV}<v_{s},v_{\Phi}<500~\text{TeV}$. The
scanned parameter intervals are chosen as follows: 
\begin{align*}
125~\text{GeV} & <m_{H_{2}}<m_{H_{3}}<4~\text{TeV},\\
78~\text{GeV} & <m_{H^{\pm}}<5~\text{TeV},\\
10~\text{GeV} & <m_{H_{0}},m_{A_{0}}<5~\text{TeV},\\
10~\text{GeV} & <m_{N_{1}}<3~\text{TeV},\\
200~\text{GeV} & <m_{Z^{\prime}}<6~\text{TeV},\\
3.32\times10^{-6} & <g_{H}<3.32.
\end{align*}
The heavy Majorana neutrino masses follow the hierarchy $m_{N_{3}}>m_{N_{2}}>m_{N_{1}}$, while the mixing angle between $Z^{\prime}$
and $Z^{\prime\prime}$ spans $-0.25<s_{\alpha}<0.25$.

\begin{figure*}[t]
\centering \begin{subfigure}[b]{0.5\textwidth} \centering\includegraphics[width=1\linewidth,keepaspectratio,height=0.27\textheight]{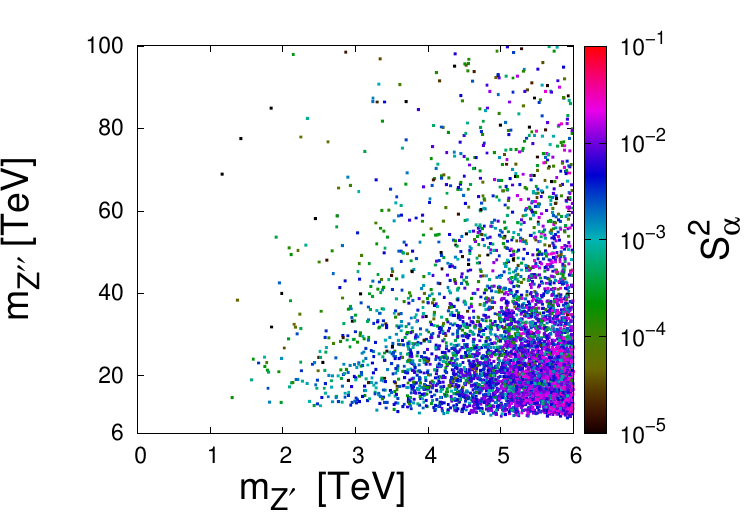}
\caption{}
\label{fig:mzpmzppgDN1} \end{subfigure}\hfill{}\begin{subfigure}[b]{0.5\textwidth}
\centering\includegraphics[width=1\linewidth,keepaspectratio,height=0.27\textheight]{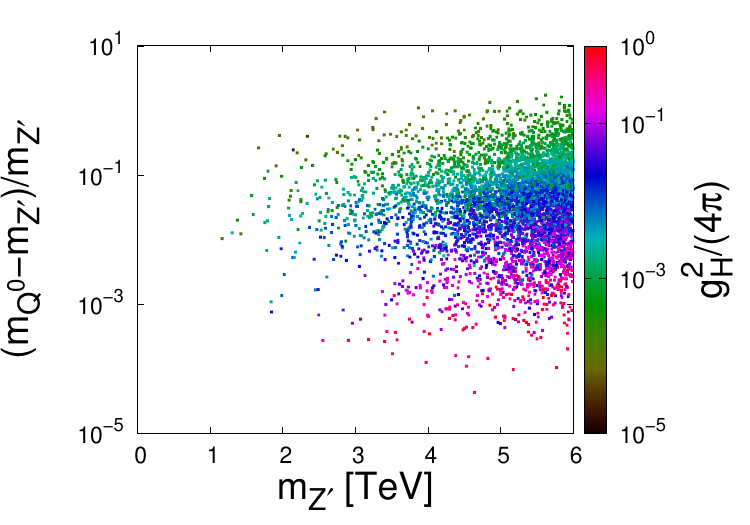}\caption{}
\label{fig:mzpmzppvpN1} \end{subfigure} \vspace{0.5em}
 \begin{subfigure}[b]{0.5\textwidth} \centering\includegraphics[width=1\linewidth,keepaspectratio,height=0.27\textheight]{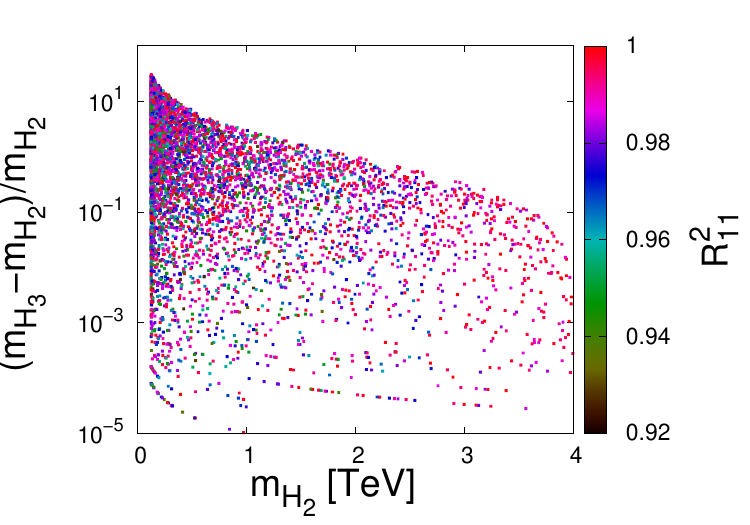}\caption{}
\label{fig:H2H3N1} \end{subfigure}\hfill{}\begin{subfigure}[b]{0.5\textwidth}
\centering\includegraphics[width=1\linewidth,keepaspectratio,height=0.27\textheight]{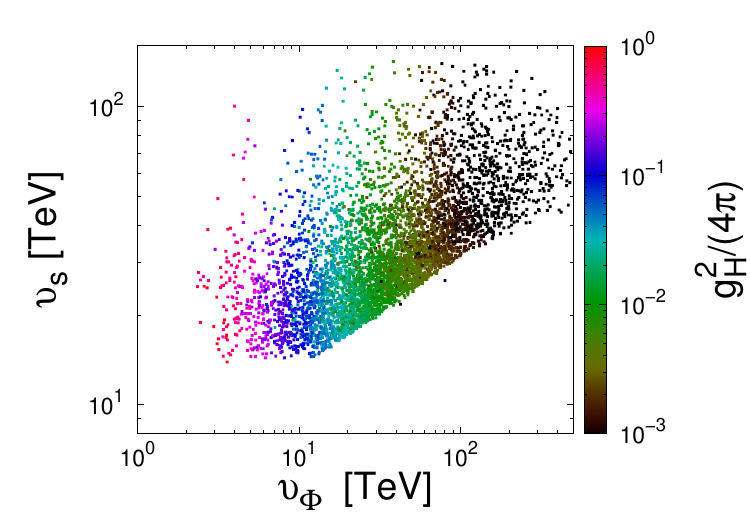}\caption{}
\label{fig:vpvsghN1} \end{subfigure}

\centering \begin{subfigure}[b]{0.5\textwidth} \centering\includegraphics[width=1\linewidth,keepaspectratio,height=0.27\textheight]{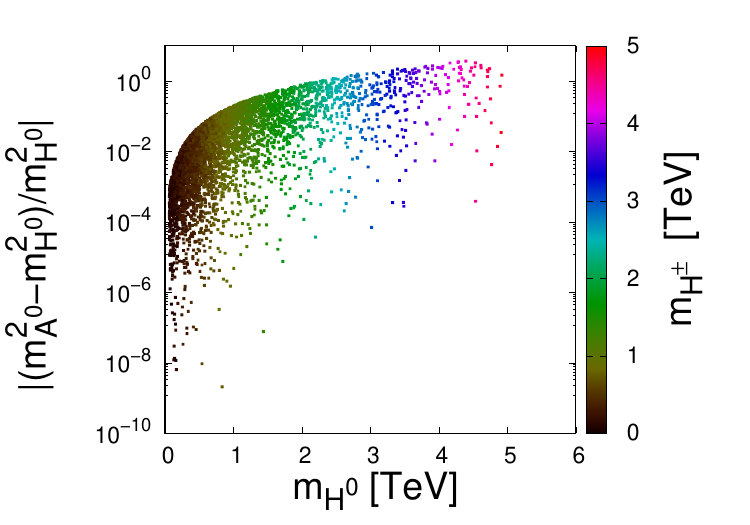}
\caption{}
\label{fig:mH0mA0mhc} \end{subfigure}\hfill{}\begin{subfigure}[b]{0.5\textwidth}
\centering\includegraphics[width=1\linewidth,keepaspectratio,height=0.27\textheight]{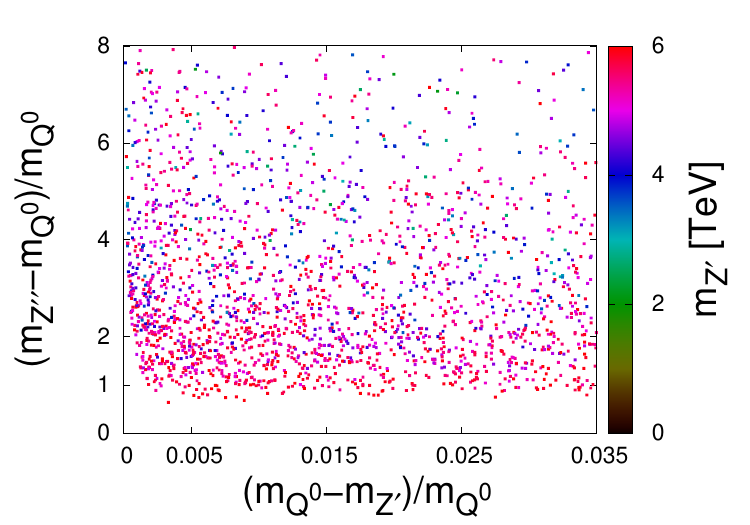}\caption{}
\label{fig:mv} \end{subfigure} \vspace{0.5em}

\caption{Correlations among the parameters of the extended gauge and scalar sectors for the {[}Majorana / Vector / Scalar{]} dark matter scenario: \textbf{(a)}~$m_{Z^{\prime\prime}}$ vs $m_{Z^{\prime}}$ colored by $s^{2}_{\alpha}$.
\textbf{(b)}~$\left(m_{Q^{0}}-m_{Z^{\prime}}\right)/m_{Z^{\prime}}$ vs $m_{Z^{\prime}}$
colored by $g^{2}_{H}/(4\pi)$. \textbf{(c)}~$\left(m_{H_{3}}-m_{H_{2}}\right)/m_{H_{2}}$
vs $m_{H_{2}}$ colored by $R^{2}_{11}$. \textbf{(d)}~Singlet scalar
VEV $v_{s}$ vs doublet VEV $v_{\Phi}$ colored by $g^{2}_{H}/(4\pi)$. \textbf{(e)} relative squared mass splitting $|(m_{A_0}^2 - m_{H_0}^2)/m_{H_0}^2|$ between the neutral inert scalars versus $m_{H_0}$, colored by the charged scalar mass $m_{H^\pm}$; and 
\textbf{(f)} relative splitting $(m_{Z^{\prime\prime}} - m_{Q^0})/m_{Q^0}$ versus $(m_{Q^0}-m_{Z'})/m_{Q^0}$, colored by $m_{Z'}$. 
Benchmark points are selected according to the theoretical and experimental
criteria described in Section~\ref{Sec:Constraints}.}
\label{Fig:Masses} 
\end{figure*}

This framework naturally accommodates three distinct DM candidates:
the lightest Majorana neutrino $N_{1}$ (MDM), the vector gauge boson
$Q^{0}$ (VDM), and the inert scalar $H_{0}$ (SDM). Their direct detection  cross sections are compared separately  with the LZ (2025) limit using the reference choice $\xi = 1$ as stated in Sec.~\ref{sec:DM_DD}.

\begin{figure*}[t]
\centering \begin{subfigure}[b]{0.47\textwidth} \centering\includegraphics[width=1\linewidth,keepaspectratio,height=0.27\textheight]{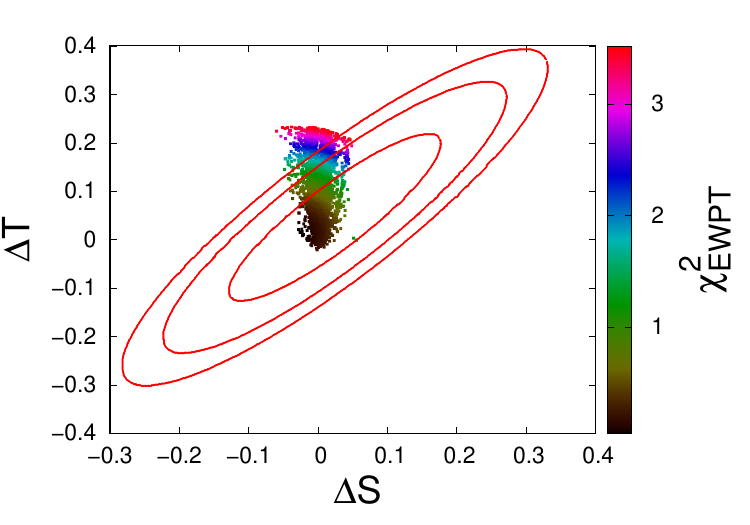}\caption{}
\label{fig:Oblique1N1} \end{subfigure}\hfill{}\begin{subfigure}[b]{0.47\textwidth}
\centering\includegraphics[width=1\linewidth,keepaspectratio,height=0.27\textheight]{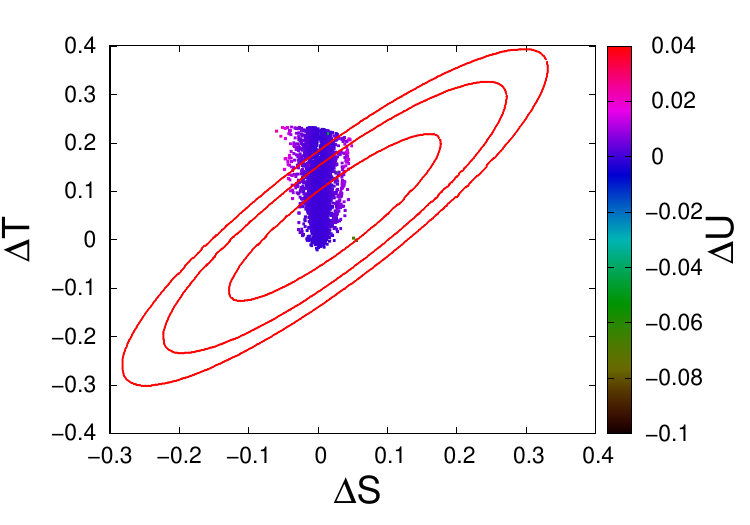}\caption{}
\label{fig:Oblique2N1} \end{subfigure} \vspace{0.5em}
 \begin{subfigure}[b]{0.47\textwidth} \centering\includegraphics[width=1\linewidth,keepaspectratio,height=0.27\textheight]{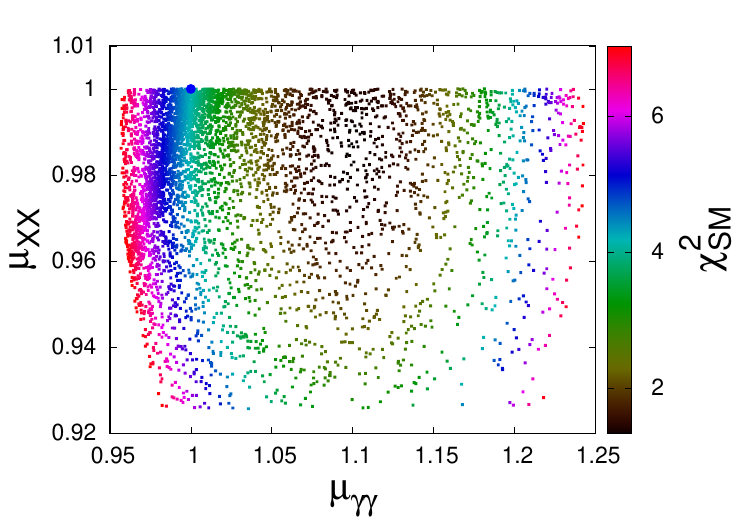}\caption{}
\label{fig:mugg} \end{subfigure}\hfill{}\begin{subfigure}[b]{0.47\textwidth}
\centering\includegraphics[width=1\linewidth,keepaspectratio,height=0.27\textheight]{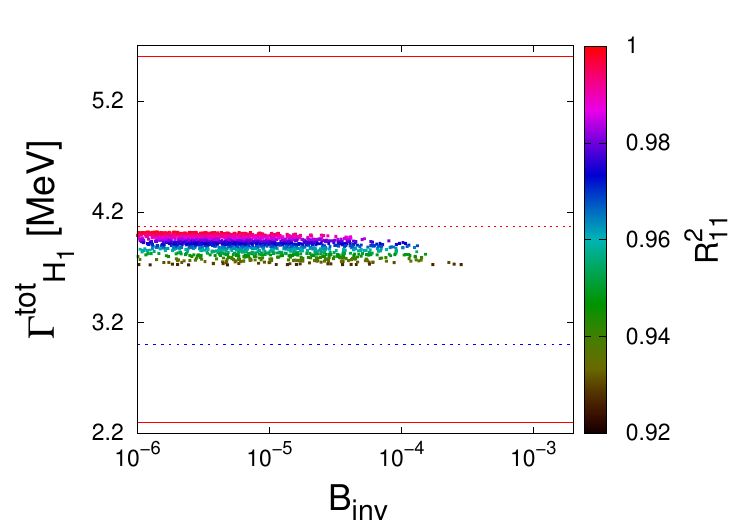}\caption{}
\label{fig:Gammat} \end{subfigure} \caption{Constraints from electroweak precision observables, Higgs signal strengths,
and the total Higgs decay width. Oblique parameters $\Delta T$ vs
$\Delta S$, overlaid with the $68\%$ (blue), $95\%$ (green), and
$99\%$ (red) CL experimental fit ellipses. The color scale of ~\textbf{(a),
(b)} represent the electroweak fit function $\chi^{2}_{\text{EWPT}}$
and $\Delta U$ respectively. \textbf{(c)}~ Higgs signal strength
modifiers $\mu_{XX}$ vs $\mu_{\gamma\gamma}$, colored by the global
chi-squared metric $\chi^{2}_{\text{SM}}$. \textbf{(d)} The total
decay width of the SM-like Higgs boson, $\Gamma^{\text{tot}}_{H_{1}}$,
as a function of the invisible branching ratio $\mathcal{B}_{\text{inv}}$,
color-coded by the $R^{2}_{11}$.}
\label{Fig:Constraints} 
\end{figure*}

To explore the interplay between the extended gauge and scalar sectors,
we perform a global parameter scan, yielding 2183 retained benchmark
points (BPs) that satisfy the applied theoretical bounds and experimental
selection criteria detailed in Sec.~\ref{Sec:Constraints}. The parametric
correlations resulting from this combined scan across all three DM
scenarios are summarized in Fig.~\ref{Fig:Masses}.

In Fig.~\ref{fig:mzpmzppgDN1} the heavy neutral gauge boson mass $m_{Z''}$ is plotted as a function of $m_{Z'}$, with the color scale indicating the square of the neutral gauge mixing angle, $s_\alpha^2$. Across the scanned parameter space, $m_{Z''}$ spans up to the scan cutoff of $100~\text{TeV}$, with larger values favored at higher $m_{Z'}$ and smaller $s_\alpha^2$, while smaller values of $s_\alpha^2$ permit a wide spread in $m_{Z''}$ at a given $m_{Z'}$. This correlation reflects the diagonalization of the neutral gauge boson mass matrix and the resulting interplay between the gauge couplings, VEVs, and mixing angle $s_\alpha$.

In Fig.~\ref{fig:mzpmzppvpN1}, the relative mass splitting $(m_{Q^{0}}-m_{Z^{\prime}})/m_{Z^{\prime}}$ is shown versus $m_{Z^{\prime}}$, with the color bar representing the gauge coupling strength $g^{2}_{H}/(4\pi)$. The fractional splitting lies primarily between $10^{-4}$ and $3\times10^{-2}$ across the scanned range of $m_{Z^{\prime}}\in[1,6]\text{TeV}$. Larger coupling strengths 
correlate with tighter relative mass degeneracies at higher $m_{Z^{\prime}}$
scales, while smaller couplings correspond to larger relative splittings.

Fig.~\ref{fig:H2H3N1} shows the relative mass splitting of the heavy
neutral scalars, $\left(m_{H_{3}}-m_{H_{2}}\right)/m_{H_{2}}$, as
a function of $m_{H_{2}}$, with the color bar representing the square
of the scalar mixing element $R^{2}_{11}$. The combined scan demonstrates
that the fractional mass splitting spans several orders of magnitude,
ranging from $10^{-5}$ up to $\sim10^{1}$ at low $m_{H_{2}}$, while
gradually tightening to $\left(m_{H_{3}}-m_{H_{2}}\right)/m_{H_{2}}\lesssim1$
as $m_{H_{2}}$ reaches $4\text{TeV}$. Across the entire parameter
space, $R^{2}_{11}$ remains tightly clustered near unity ($0.92\le R^{2}_{11}\le1.0$), indicating that the $125\text{GeV}$ state ($H_{1}$) retains strictly SM-like couplings to gauge bosons and fermions. This suppression of mixing with the heavy scalar sector is essential for satisfying the combined constraints from LHC Higgs signal strengths and electroweak precision observables.

Fig.~\ref{fig:vpvsghN1} represents the scalar singlet VEV
$v_{s}$ plotted against the scalar doublet VEV $v_{\Phi}$, with
the color bar again indicating $g^{2}_{H}/(4\pi)$. The viable BPs span $v_s \sim 8\text{-}15~\text{TeV}$ and $v_\Phi \sim 1\text{-}20~\text{TeV}$ in the displayed region, with larger $g_H^2/(4\pi)$ favoring smaller $v_\Phi$ and smaller couplings extending to larger $v_\Phi$. In Fig.~\ref{fig:mH0mA0mhc}, the relative squared mass splitting $|(m_{A_0}^2 - m_{H_0}^2)/m_{H_0}^2|$ between the neutral inert scalars is shown as a function of $m_{H_0}$, with the color bar indicating the charged scalar mass $m_{H^\pm}$. The fractional squared splitting spans several orders of magnitude, from $\sim 10^{-10}$ to $\mathcal{O}(1\text{-}10)$ at lower $m_{H_0}$, while its lower boundary rises toward larger $m_{H_0}$, with larger $m_{H^\pm}$ values becoming increasingly prevalent in the multi-TeV regime. Small splittings between $H_0$ and $A_0$ correspond to suppressed $\lambda_5$, which naturally helps satisfy one-loop neutrino mass constraints alongside LFV bounds.

Finally, Fig.~\ref{fig:mv} shows the relative mass splittings $(m_{Z^{\prime\prime}}-m_{Q^0})/m_{Q^0}$ versus $(m_{Q^0}-m_{Z'})/m_{Q^0}$, with the points colored according to $m_{Z'}$. For the retained BPs, $(m_{Q^0}-m_{Z'})/m_{Q^0}$ remains small, reaching at most $\sim 0.035$, indicating a close mass degeneracy between $Q^0$ and $Z'$. In contrast, $(m_{Z^{\prime\prime}}-m_{Q^0})/m_{Q^0}$ ranges from $\sim 0.8$ to $\sim 8$ over the scanned $m_{Z'}$ range, extending up to $6~\text{TeV}$. This reflects the sizable mass separation between the lighter $(Z',Q^0)$ states and the heavier $Z^{\prime\prime}$ eigenstate.

To evaluate how precision constraints restrict the parameter space,
Fig.~\ref{Fig:Constraints}, illustrates the distribution of BPs
combining the three dark matter scenarios (MDM, VDM, and SDM) in light
of electroweak precision observables and Higgs signal strength measurements.
Fig.~\ref{fig:Oblique1N1}, and Fig.~\ref{fig:Oblique2N1} show
the constraints imposed by the oblique parameters $\Delta S$ and
$\Delta T$, overlaid on the $68\%$, $95\%$, and $99\%$ C.L.\ ellipses
derived from global EWPT fits, with the color palettes representing
$\chi^{2}_{\text{EWPT}}$ in Fig.~\ref{fig:Oblique1N1} and $\Delta U$
Fig.~\ref{fig:Oblique2N1}. The BPs are clustered near $\Delta S=0$,
although the plotted distribution extends beyond the $95\%$ C.L.\ ellipse.
The dispersion is primarily vertical, spanning $0.00\lesssim\Delta T\lesssim0.24$, which reflects the sensitivity of $\Delta T$ to mass splittings among the inert doublet components ($H_{0}$, $A_{0}$, $H^{\pm}$) and the extended CP-even scalar sector. Conversely, $\Delta S$ remains
tightly clustered near zero ($-0.05\lesssim\Delta S\lesssim0.08$)
due to its relatively weak logarithmic dependence on mass splittings,
while $\Delta U\approx0$ remains strongly suppressed.

In Fig.~\ref{fig:mugg}, the correlation between the Higgs signal
strength modifier into standard tree-level modes $\mu_{XX}$ ($X=b,\tau,\mu,W,Z$) and the diphoton channel $\mu_{\gamma\gamma}$ is shown, colored by the global $\chi^{2}_{\text{SM}}$. The BPs concentrate near $\mu_{XX}\sim0.93$\,\,-\,\,$1.00$, enforcing the SM-like Higgs alignment limit ($R_{11}\approx1$), whereas $\mu_{\gamma\gamma}$ extends from $0.95$ up to $\approx1.25$ due to additional loop contributions from the charged scalar $H^{\pm}$. The SM prediction $(\mu_{\gamma\gamma},\mu_{XX})=(1,1)$ is indicated by the orange benchmark marker, and all valid BPs yield $\chi^{2}_{\text{SM}}\le4$,
consistent with the applied Higgs signal strength constraints.

The correlation between the total decay width of the SM-like Higgs
boson ($\Gamma^{\text{tot}}_{H_{1}}$), its invisible branching ratio
($\mathcal{B}_{\text{inv}}$), and the scalar mixing matrix element
($R^{2}_{11}$) is presented in Fig.~\ref{fig:Gammat}. Across the
entire viable parameter space, the invisible branching ratio remains
tiny ($\mathcal{B}_{\text{inv}}\lesssim2\times10^{-3}$), well below
the experimental upper bound. The total decay width is tightly constrained
within the narrow interval $3.70~\text{MeV}\lesssim\Gamma^{\text{tot}}_{H_{1}}\lesssim4.02~\text{MeV}$,
staying comfortably inside the experimental bounds (represented by
the horizontal lines) and slightly suppressed relative to the SM expectation
($\Gamma_{\text{SM}}\approx4.02~\text{MeV}$) due to non-zero scalar
mixing ($0.92\le R^{2}_{11}\le1.0$).

\begin{figure*}[t]
\begin{subfigure}[b]{0.47\textwidth} \includegraphics[width=1\linewidth,keepaspectratio,height=0.27\textheight]{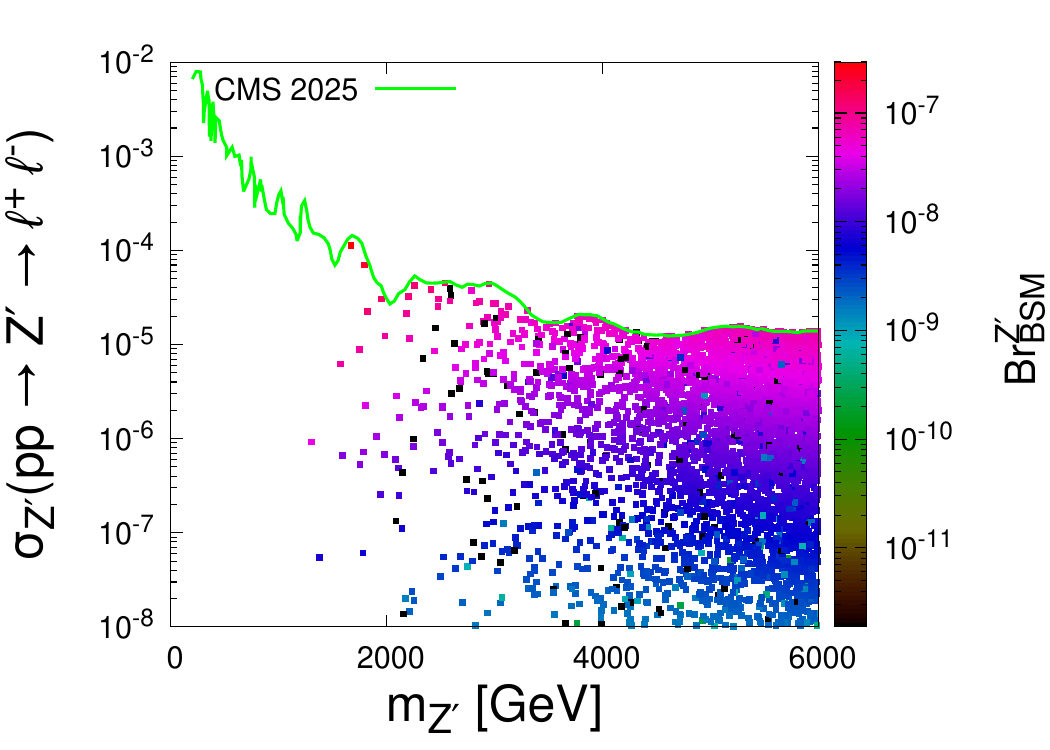}
\caption{}
\label{fig:SigZp} \end{subfigure}\hfill{}\begin{subfigure}[b]{0.47\textwidth}
\includegraphics[width=1\linewidth,keepaspectratio,height=0.27\textheight]{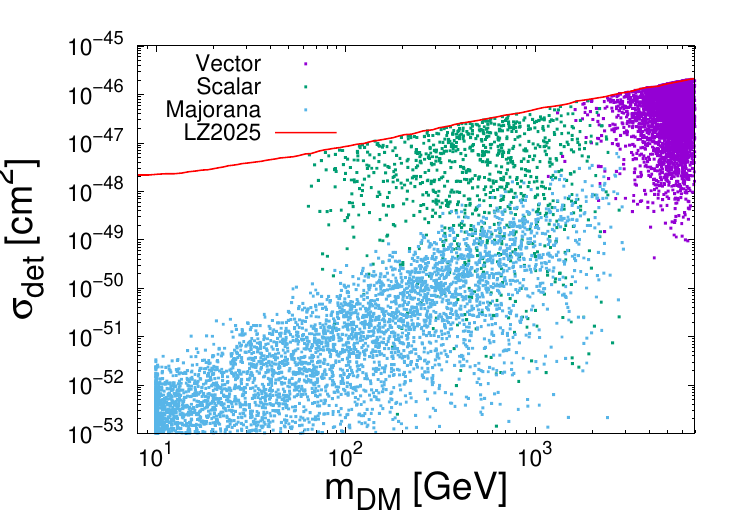}
\caption{}
\label{fig:SDD} \end{subfigure} \vspace{0.5em}
 \caption{Experimental bounds on the heavy gauge sector and direct dark matter detection constraints. \textbf{(a)}~Dilepton production cross-section ratio $\sigma(pp\to Z^{\prime}\to \ell^{+}\ell^{-})$ as a function of the heavy
gauge boson mass $m_{Z^{\prime}}$, compared with CMS experimental limits~\cite{CMS:2026xoc}. \textbf{(b)}~Spin-independent dark matter--nucleon scattering cross section $\sigma_{\text{det}}$ versus DM mass for the three DM  candidates: Majorana DM ($N_{1}$), vector DM ($Q^{0}$), and scalar DM ($H_{0}$). Benchmark points are evaluated against the upper limit from the LZ experiment~\cite{LZ:2024zvo} (solid red curve).}
\label{Fig:SDDZp} 
\end{figure*}

In Fig.~\ref{Fig:SDDZp}, we examine the phenomenological viability
of the  parameter space by evaluating the limits arising
from direct heavy gauge boson searches at the LHC~\cite{ATLAS:2019erb,CMS:2026xoc}
alongside spin-independent DM DD experiments~\cite{LZ:2024zvo}.
Fig.~\ref{fig:SigZp} illustrate the
dilepton production cross-section ratio $\sigma(pp\to Z^{\prime}\to \ell^{+}\ell^{-})$ for the heavy neutral gauge states $Z^{\prime}$ as a function of their respective masses, $m_{Z^{\prime}}$. These theoretical predictions are compared against the CMS 13~TeV upper bounds obtained from high-mass dilepton resonance channels ($pp\to Z^{\prime}\to\ell^{+}\ell^{-}$,
where $\ell=e,\mu$).

The physical mass eigenstates $Z^{\prime}$ and $Z^{\prime\prime}$ arise from a mixture of the extra $U(1)$ gauge field combination 
$(B^1-B^2)/\sqrt{2}$ and the third $SU(2)_H$ component $Q^3$. The suppressed
$Z^{\prime}$ dilepton production rates observed across a significant portion of the scanned parameter space can be understood  from  small neutral mixing angles ($|s_{\alpha}|\ll1$), which  suppress its couplings to SM fermions. From Fig.~\ref{fig:SigZp},
the CMS bounds enforce a lower limit of $m_{Z^{\prime}}\gtrsim2\text{TeV}$
for points characterized by moderate gauge mixing. Conversely, in
regions where the mixing angle is heavily suppressed, the $Z^{\prime}$ state
decoupling occurs naturally, allowing lighter mass eigenvalues to
remain viable. For the the neutral gauge mixing angle is bounded by $|s_\alpha| < 0.25$, the
heavier neutral gauge boson $Z^{\prime\prime}$ retains an $\mathcal{O}(1)$
coupling factor ($c_\alpha \approx 1$) to SM fermions, unlike $Z'$ whose
fermionic couplings are suppressed by $s_\alpha$. Consequently, lighter
$Z^{\prime\prime}$ masses are tightly excluded by CMS high-mass dilepton
resonance searches, leaving almost no viable points in the mass range directly
constrained by the CMS bound and pushing surviving parameter points toward
heavier $m_{Z^{\prime\prime}}$ scales (where dilepton rates fall and competing
channels such as $Z^{\prime\prime} \to V H_i$ can further reduce the dilepton
branching ratio when kinematically open). These results motivate further studies of the $Z^{\prime\prime}$ discovery potential at  future high-energy colliders.

Concerning dark matter, the SI DD nucleon scattering cross section,
$\sigma_{\text{det}}$, is evaluated across all three distinct DM
candidates ($N_{1}$, $H_{0}$, and $Q^{0}$) and compared against
the latest experimental bounds from the LZ (2025) collaboration~\cite{LZ:2024zvo},
as displayed in Fig.~\ref{fig:SDD}: 
\begin{enumerate}
\item {\bf Majorana Fermion DM} ($N_{1}$): For the lightest Majorana singlet
neutrino (blue points), the elastic scattering off nuclei proceeds
via $t$-channel exchange of neutral scalar fields driven by the scalar
singlet mixing elements $R_{i1}$. Because the underlying Yukawa couplings
are strongly constrained by low-energy LFV processes, most notably
$\mu\to e\gamma$, the resulting tree-level scattering amplitudes
are naturally suppressed. The corresponding cross sections cluster
in the range $\sigma_{\text{det}}\sim10^{-52}\text{cm}^{2}\; \text{to}\; 10^{-47}\text{cm}^{2}$, remaining comfortably below the current LZ exclusion curve and future xenon-based detection limits.
\item {\bf Scalar DM} ($H_{0}$): In the case of the inert scalar doublet state (orange points), direct detection is dominated by tree-level $t$-channel Higgs mediation via the trilinear portal couplings $\lambda_{H_{0}H_{0}H_{i}}$. Parameter viability relative to the LZ bound is restricted to two primary regimes: the narrow Higgs pole resonance region ($m_{\text{DM}}\approx m_{H_{1}}/2$), and the high mass regime ($m_{\text{DM}}\gtrsim500\text{GeV}$) where efficient co-annihilation channels ($H_{0}A_{0}\to Z$ and $H_{0}H^{\pm}\to W^{\pm}$) deplete the thermal relic density without requiring large scalar portal couplings.
\item {\bf Hidden Vector DM} ($Q^{0}$): The hidden gauge vector candidate (magenta points) resides predominantly in the multi-TeV regime ($m_{\text{DM}}\sim1\,\,\text{-}\,3\;\text{TeV}$) with its  direct detection cross sections spanning $10^{-50}\text{cm}^{2}\text{to}10^{-46}\text{cm}^{2}$. Since $Q^{0}$ interacts with SM fields 
through the scalar-portal couplings $\lambda_{H_iQ^0\bar Q^0}\propto g^{2}_{H}v_{\Phi}R_{i3}$ (and, in principle, through $Z'/Z''$ exchange), the $Q^{0}N$ and $\bar{Q}^{0}N$ cross sections receive scalar and gauge boson contributions. Assuming no particle antiparticle asymmetry, the effective cross section is their average, and satisfying the LZ (2025) constraints requires either heavy vector masses or near unit scalar mixing ($R_{11}\approx1$). Under these conditions, nuclear recoil rates are kept below experimental thresholds while the required thermal relic abundance could in principle be maintained via efficient hidden-sector annihilation channels, $Q^{0}\bar{Q}^{0}\to Z^{\prime}Z^{\prime},H_{i}H_{j},f\bar{f}$, which will be quantified in a future work together with multi component DM effects. 
\end{enumerate}

\section{Conclusion~\label{Sec:Conclusion}}

In this work, we have investigated an extended gauge symmetry framework
based on $SU(2)_{L}\times SU(2)_{H}\times U(1)_{1}\times U(1)_{2}$,
designed to jointly account for the origin of tiny active neutrino
masses and the nature of DM. By enlarging the particle content to
include an inert electroweak scalar doublet ($\eta$), a hidden $SU(2)_{H}$
scalar doublet ($\Phi$), a complex scalar singlet ($S$), and three
heavy Majorana singlet fermions ($N_{i}$), the model implements a
scotogenic mechanism where active neutrino masses are successfully
generated at the one-loop level under an exact $Z_{2}$ symmetry.

Depending on the mass hierarchy and the chosen sector parameters,
the framework naturally accommodates three distinct dark matter candidates:
the lightest Majorana fermion ($N_{1}$), the inert scalar ($H_{0}$),
and the hidden vector gauge boson ($Q^{0}$). Through a comprehensive
global parameter scan, the reported points were selected using theoretical consistency requirements, including perturbative unitarity and vacuum stability as well as strict observational constraints from electroweak precision tests (oblique parameters $S,T,U$), LHC Higgs signal strengths and total decay width limits, lepton flavor violation bounds, and direct DM detection limits from the LZ experiment.

Our numerical analysis illustrate  that the model is compatible with the 
current experimental bounds under the stated assumptions while maintaining a predictive and tightly constrained parameter space. Specifically, the SM-like Higgs couplings remain well aligned with standard expectations, and the heavy neutral gauge bosons ($Z^{\prime}$ and $Z^{\prime\prime}$) are compared with the current LHC dilepton resonance searches while offering exciting prospects for future high energy collider investigations.

\vskip 1truecm \textbf{Acknowledgements}: A.A. is funded by the University
of Sharjah via the HEP RG operational grant. The work of M.O.K. is
supported by SAFE (Supporting At-Risk Researchers with Fellowships
in Europe project), which is funded by the European Union (EU) under
the Grant Agreement ref. 101148426. S.N. is supported by the United
Arab Emirates University (UAEU) under UPAR Grant No. 12S162. AI tools were used to polish the language in the text.

\appendix

\section{Anomaly Cancellation and Charge Values~\label{app:anom}}

Before symmetry breaking, each $N_{i}$ is a right-handed chiral singlet
whose Majorana mass arises dynamically once $S$ acquires a vacuum
expectation value. Let $q_{f}$ and $t_{f}$ denote the $U(1)_{1}$
and $U(1)_{2}$ charges of field $f$ (with $x_{f}$ representing
either charge), and let $u,d,e,N$ represent the right-handed fermions.
Imposing gauge invariance on the Yukawa terms in Eq.~\eqref{eq:Yuk:1}
and on $(H^{\dagger}\eta)^{2}$ leads to: 
\begin{align}
x_{e} & =x_{L}-x_{H}, & x_{d} & =x_{Q}-x_{H},\nonumber \\
x_{u} & =x_{Q}+x_{H}, & x_{N} & =x_{L}+x_{\eta},\nonumber \\
x_{S} & =-2x_{N}, & x_{\eta} & =x_{H}.\label{eq:charge-relations}
\end{align}

Gauge and mixed gravitational anomalies receive contributions exclusively
from chiral fermions, where right-handed fields enter with a negative
sign when expressed as left-handed conjugates. Omitting a common normalization
factor, the conditions for each generation and for each charge $x\in\{q,t\}$
are: 
\begin{align}
\mathcal{A}_{33}(x) & =2x_{Q}-x_{u}-x_{d}=0,\nonumber \\
\mathcal{A}_{22}(x) & =3x_{Q}+x_{L}=0,\nonumber \\
\mathcal{A}_{\mathrm{grav}}(x) & =6x_{Q}+2x_{L}-3x_{u}-3x_{d}\nonumber \\
 & \quad-x_{e}-x_{N}=0,\nonumber \\
\mathcal{A}_{3}(x) & =6x^{3}_{Q}+2x^{3}_{L}-3x^{3}_{u}-3x^{3}_{d}\nonumber \\
 & \quad-x^{3}_{e}-x^{3}_{N}=0.\label{eq:pure-anomalies}
\end{align}
Here $33$ and $22$ label the $SU(3)_{C}\,SU(3)_{C}\,U(1)$ and $SU(2)_{L}\,SU(2)_{L}\,U(1)$
anomalies, respectively. The two mixed Abelian conditions are given
by: 
\begin{align}
\mathcal{A}_{112} & =6q^{2}_{Q}t_{Q}+2q^{2}_{L}t_{L}\nonumber \\
 & \quad-3q^{2}_{u}t_{u}-3q^{2}_{d}t_{d}\nonumber \\
 & \quad-q^{2}_{e}t_{e}-q^{2}_{N}t_{N}=0,\nonumber \\
\mathcal{A}_{122} & =6q_{Q}t^{2}_{Q}+2q_{L}t^{2}_{L}\nonumber \\
 & \quad-3q_{u}t^{2}_{u}-3q_{d}t^{2}_{d}\nonumber \\
 & \quad-q_{e}t^{2}_{e}-q_{N}t^{2}_{N}=0.\label{eq:mixed-anomalies}
\end{align}
Since all fermions are neutral under $SU(2)_{H}$ (singlets), no chiral
$SU(2)_{H}$ anomalies arise. The $SU(2)_{L}$ Witten anomaly is similarly
absent, given that each generation comprises four left-handed doublets
(including the three colors of $Q_{L}$).

The unbroken hypercharge generator may initially be written as 
\begin{equation}
Y=aT^{3}_{H}+bN_{1}+cN_{2},\label{eq:ansatz-1}
\end{equation}
where $N_{1,2}$ carry charges $q_{f}$ and $t_{f}$. Adopting the
charge basis of the main text, we impose $t_{S}=-q_{S}$ and $t_{\Phi}=-q_{\Phi}$
to ensure that both scalar vacuum expectation values are neutral under
the targeted hypercharge. For $q_{S}\neq0$, the neutrality condition
for $\langle S\rangle$ in Eq.~\eqref{eq:ansatz-1} yields $b=c$.
Given that the lower component of $\Phi$ has $T^{3}_{H}=-1/2$, its
neutrality requires $-a/2+b(q_{\Phi}+t_{\Phi})=0$, which implies
$a=0$. Finally, normalizing the SM hypercharges fixes $b=c=1$, and
consequently: 
\begin{equation}
Y=N_{1}+N_{2},\qquad Y_{f}=q_{f}+t_{f}.\label{eq:hypercharge-sum}
\end{equation}
In particular, the hypercharges are given by $Y_{Q}=1/6$, $Y_{u}=2/3$,
$Y_{d}=-1/3$, $Y_{L}=-1/2$, $Y_{e}=-1$, $Y_{N}=0$, and $Y_{H}=Y_{\eta}=1/2$.

Combining Eq.~\eqref{eq:charge-relations} with the anomaly cancellation
condition $\mathcal{A}_{22}(q)=0$ determines the $U(1)_{1}$ charges
in terms of $q_{\eta}$ and $q_{S}$: 
\begin{align}
q_{H} & =q_{\eta}, & q_{N} & =-\tfrac{1}{2}q_{S},\nonumber \\
q_{L} & =-q_{\eta}-\tfrac{1}{2}q_{S}, & q_{Q} & =\tfrac{1}{3}q_{\eta}+\tfrac{1}{6}q_{S},\nonumber \\
q_{u} & =\tfrac{4}{3}q_{\eta}+\tfrac{1}{6}q_{S}, & q_{d} & =-\tfrac{2}{3}q_{\eta}+\tfrac{1}{6}q_{S},\nonumber \\
q_{e} & =-2q_{\eta}-\tfrac{1}{2}q_{S}.\label{eq:general-charges}
\end{align}
The second set is given by $t_{f}=Y_{f}-q_{f}$, with $t_{S}=-q_{S}$
and $t_{H}=t_{\eta}=1/2-q_{\eta}$. Meanwhile, the scalar charge $q_{\Phi}=-t_{\Phi}$
remains independent. Consequently, the complete charge assignment
is parameterized solely by $q_{S}$, $q_{\eta}$, and $q_{\Phi}$,
free of any inverse powers of $q_{S}$.

Anomaly cancellation is evaluated transparently without substituting
individual charges into Eqs.~\eqref{eq:pure-anomalies} and \eqref{eq:mixed-anomalies}
by defining $X=B-L$ for the fermions, with $X_{Q,u,d}=1/3$ and $X_{L,e,N}=-1$.
In this basis, Eq.~\eqref{eq:general-charges} equivalently reads:
\begin{align}
q_{f} & =2q_{\eta}Y_{f}+\tfrac{1}{2}q_{S}X_{f},\nonumber \\
t_{f} & =(1-2q_{\eta})Y_{f}-\tfrac{1}{2}q_{S}X_{f}.\label{eq:charge-basis}
\end{align}
Using the chiral multiplicities $\sigma_{f}=(6,2,-3,-3,-1,-1)$ for
$f\in\{Q,L,u,d,e,N\}$, and assuming a single right-handed neutrino
$N_{i}$ per generation, direct summation yields: 
\begin{align}
\sum_{f}\sigma_{f}Y_{f} & =\sum_{f}\sigma_{f}X_{f}=0,\nonumber \\
\sum_{f}\sigma_{f}Y^{3}_{f} & =\sum_{f}\sigma_{f}Y^{2}_{f}X_{f}=0,\nonumber \\
\sum_{f}\sigma_{f}Y_{f}X^{2}_{f} & =\sum_{f}\sigma_{f}X^{3}_{f}=0.\label{eq:anomaly-traces}
\end{align}
Furthermore, the $[SU(3)_{C}]^{2}$ and $[SU(2)_{L}]^{2}$ traces
vanish independently for both $Y$ and $X$. Because $q$ and $t$
are linear combinations of these generators, Eqs.~\eqref{eq:anomaly-traces}
guarantee the complete cancellation of both pure cubic anomalies,
both mixed Abelian anomalies, and both mixed gravitational anomalies
for arbitrary values of $q_{\eta}$ and $q_{S}$.

For the benchmark scenario used in the main text, setting $q_{S}=-1$,
$q_{\eta}=1/4$, and $q_{\Phi}=1/4$ yields the following charge pairs
$(q_{f},t_{f})$: 
\begin{align}
(q_{Q},t_{Q}) & =(-\tfrac{1}{12},\tfrac{1}{4}),\nonumber \\
(q_{u},t_{u}) & =(\tfrac{1}{6},\tfrac{1}{2}),\nonumber \\
(q_{d},t_{d}) & =(-\tfrac{1}{3},0),\nonumber \\
(q_{L},t_{L}) & =(\tfrac{1}{4},-\tfrac{3}{4}),\nonumber \\
(q_{e},t_{e}) & =(0,-1),\nonumber \\
(q_{N},t_{N}) & =(\tfrac{1}{2},-\tfrac{1}{2}),\nonumber \\
(q_{H},t_{H}) & =(q_{\eta},t_{\eta})=(\tfrac{1}{4},\tfrac{1}{4}),\nonumber \\
(q_{S},t_{S}) & =(-1,1),\nonumber \\
(q_{\Phi},t_{\Phi}) & =(\tfrac{1}{4},-\tfrac{1}{4}).\label{eq:charges-1}
\end{align}
Specifically, the two mixed Abelian sums evaluate to:
\begin{align}
\mathcal{A}_{112} & =-\tfrac{3}{32}+\tfrac{1}{96}-\tfrac{1}{24}+\tfrac{1}{8}=0,\nonumber \\
\mathcal{A}_{122} & =\tfrac{9}{32}-\tfrac{1}{32}-\tfrac{1}{8}-\tfrac{1}{8}=0.
\end{align}
The remaining conditions vanish identically via the same charge-basis
argument.

\section{Vacuum Stability \& Unitarity Matrices~\label{app:Unitarity}}

The matrix $M_0$ that relevant to the vacuum stability constraints in Eq.~\ref{eq:VStab} is
\begin{align}
\left(\begin{array}{ccccccccccc}
\lambda_H & \bar{\omega}_2 & \bar{\omega}_1 & \bar{\lambda}_+ & 0 & \bar{\omega}_2 & \bar{\omega}_1 & \bar{\lambda}_- & 0 & \bar{\lambda}_3 & \bar{\omega}_1\\
\bar{\omega}_2 & \lambda_S & \bar{\omega}_4 & \bar{\omega}_5 & \bar{\omega}_2 & 0 & \bar{\omega}_4 & \bar{\omega}_5 & \bar{\omega}_2 & \bar{\omega}_5 & \bar{\omega}_4\\
\bar{\omega}_1 & \bar{\omega}_4 & \lambda_\Phi & \bar{\omega}_3 & \bar{\omega}_1 & \bar{\omega}_4 & 0 & \bar{\omega}_3 & \bar{\omega}_1 & \bar{\omega}_3 & 0\\
\bar{\lambda}_+ & \bar{\omega}_5 & \bar{\omega}_3 & \lambda_\eta & \bar{\lambda}_- & \bar{\omega}_5 & \bar{\omega}_3 & 0 & \bar{\lambda}_3 & 0 & \bar{\omega}_3\\
0 & \bar{\omega}_2 & \bar{\omega}_1 & \bar{\lambda}_- & \lambda_H & \bar{\omega}_2 & \bar{\omega}_1 & \bar{\lambda}_+ & 0 & \bar{\lambda}_3 & \bar{\omega}_1\\
\bar{\omega}_2 & 0 & \bar{\omega}_4 & \bar{\omega}_5 & \bar{\omega}_2 & \lambda_S & \bar{\omega}_4 & \bar{\omega}_5 & \bar{\omega}_2 & \bar{\omega}_5 & \bar{\omega}_4\\
\bar{\omega}_1 & \bar{\omega}_4 & 0 & \bar{\omega}_3 & \bar{\omega}_1 & \bar{\omega}_4 & \lambda_\Phi & \bar{\omega}_3 & \bar{\omega}_1 & \bar{\omega}_3 & 0\\
\bar{\lambda}_- & \bar{\omega}_5 & \bar{\omega}_3 & 0 & \bar{\lambda}_+ & \bar{\omega}_5 & \bar{\omega}_3 & \lambda_\eta & \bar{\lambda}_3 & 0 & \bar{\omega}_3\\
0 & \bar{\omega}_2 & \bar{\omega}_1 & \bar{\lambda}_3 & 0 & \bar{\omega}_2 & \bar{\omega}_1 & \bar{\lambda}_3 & \frac{2}{3}\lambda_H & \bar{\lambda}_{34} & \bar{\omega}_1\\
\bar{\lambda}_3 & \bar{\omega}_5 & \bar{\omega}_3 & 0 & \bar{\lambda}_3 & \bar{\omega}_5 & \bar{\omega}_3 & 0 & \bar{\lambda}_{34} & \frac{2}{3}\lambda_\eta & \bar{\omega}_3\\
\bar{\omega}_1 & \bar{\omega}_4 & 0 & \bar{\omega}_3 & \bar{\omega}_1 & \bar{\omega}_4 & 0 & \bar{\omega}_3 & \bar{\omega}_1 & \bar{\omega}_3 & \frac{2}{3}\lambda_\Phi
\end{array}\right),
\end{align}
with $\bar{X}=\min(0,X)$.

In this appendix, we present the relevant scattering amplitude matrices
and their corresponding eigenvalues for the unitarity analysis. Although
the full neutral scalar scattering matrix is $41\times41$-dimensional,
it block-diagonalizes into three independent submatrices: 
\begin{equation}
\mathcal{M}^{N}_{41\times41}=\left(\begin{array}{ccc}
\left(\mathcal{M}^{N}_{1}\right)_{11\times11} & 0 & 0\\
0 & \left(\mathcal{M}^{N}_{2}\right)_{22\times22} & 0\\
0 & 0 & \left(\mathcal{M}^{N}_{3}\right)_{8\times8}
\end{array}\right),
\end{equation}

\begin{widetext}
where the neutral scalar submatrix $\mathcal{M}^{N}_{1}$ expressed
in the basis 
\begin{equation}
\left\{ \frac{h_{1}h_{1}}{\sqrt{2}},\frac{h_{2}h_{2}}{\sqrt{2}},\frac{h_{3}h_{3}}{\sqrt{2}},\frac{h_{4}h_{4}}{\sqrt{2}},\frac{a_{1}a_{1}}{\sqrt{2}},\frac{a_{2}a_{2}}{\sqrt{2}},\frac{a_{3}a_{3}}{\sqrt{2}},\frac{a_{4}a_{4}}{\sqrt{2}},h^{+}_{1}h^{-}_{1},h^{+}_{2}h^{-}_{2},\phi_{0}\phi^{*}_{0}\right\} 
\end{equation}
is given by 
\begin{equation}
\mathcal{M}^{N}_{1}=\left(\begin{array}{ccccccccccc}
\frac{1}{2}\lambda_{H}\,\, & \frac{1}{2}\omega_{2}\,\, & \frac{1}{2}\omega_{1}\,\, & \frac{1}{2}\lambda_{+}\,\, & \frac{1}{6}\lambda_{H}\,\, & \frac{1}{2}\omega_{2}\,\, & \frac{1}{2}\omega_{1}\,\, & \frac{1}{2}\lambda_{-}\,\, & \frac{1}{3\sqrt{2}}\lambda_{H} & \frac{1}{\sqrt{2}}\lambda_{3} & \frac{1}{\sqrt{2}}\omega_{1}\\
\frac{1}{2}\omega_{2} & \frac{1}{2}\lambda_{S} & \frac{1}{2}\omega_{4} & \frac{1}{2}\omega_{5} & \frac{1}{2}\omega_{2} & \frac{1}{6}\lambda_{S} & \frac{1}{2}\omega_{4} & \frac{1}{2}\omega_{5} & \frac{1}{\sqrt{2}}\omega_{2} & \frac{1}{\sqrt{2}}\omega_{5} & \frac{1}{\sqrt{2}}\omega_{4}\\
\frac{1}{2}\omega_{1} & \frac{1}{2}\omega_{4} & \frac{1}{2}\lambda_{\Phi} & \frac{1}{2}\omega_{3} & \frac{1}{2}\omega_{1} & \frac{1}{2}\omega_{4} & \frac{1}{6}\lambda_{\Phi} & \frac{1}{2}\omega_{3} & \frac{1}{\sqrt{2}}\omega_{1} & \frac{1}{\sqrt{2}}\omega_{3} & \frac{1}{3\sqrt{2}}\lambda_{\Phi}\\
\frac{1}{2}\lambda_{+} & \frac{1}{2}\omega_{5} & \frac{1}{2}\omega_{3} & \frac{1}{2}\lambda_{\eta} & \frac{1}{2}\lambda_{-} & \frac{1}{2}\omega_{5} & \frac{1}{2}\omega_{3} & \frac{1}{6}\lambda_{\eta} & \frac{1}{\sqrt{2}}\lambda_{3} & \frac{1}{3}\lambda_{\eta} & \frac{1}{\sqrt{2}}\omega_{3}\\
\frac{1}{6}\lambda_{H} & \frac{1}{2}\omega_{2} & \frac{1}{2}\omega_{1} & \frac{1}{2}\lambda_{-} & \frac{1}{2}\lambda_{H} & \frac{1}{2}\omega_{2} & \frac{1}{2}\omega_{1} & \frac{1}{2}\lambda_{+} & \frac{1}{3\sqrt{2}}\lambda_{H} & \frac{1}{\sqrt{2}}\lambda_{3} & \frac{1}{\sqrt{2}}\omega_{1}\\
\frac{1}{2}\omega_{2} & \frac{1}{6}\lambda_{S} & \frac{1}{2}\omega_{4} & \frac{1}{2}\omega_{5} & \frac{1}{2}\omega_{2} & \frac{1}{2}\lambda_{S} & \frac{1}{2}\omega_{4} & \frac{1}{2}\omega_{5} & \frac{1}{\sqrt{2}}\omega_{2} & \frac{1}{\sqrt{2}}\omega_{5} & \frac{1}{\sqrt{2}}\omega_{4}\\
\frac{1}{2}\omega_{1} & \frac{1}{2}\omega_{4} & \frac{1}{6}\lambda_{\Phi} & \frac{1}{2}\omega_{3} & \frac{1}{2}\omega_{1} & \frac{1}{2}\omega_{4} & \frac{1}{2}\lambda_{\Phi} & \frac{1}{2}\omega_{3} & \frac{1}{\sqrt{2}}\omega_{1} & \frac{1}{\sqrt{2}}\omega_{3} & \frac{1}{3\sqrt{2}}\lambda_{\Phi}\\
\frac{1}{2}\lambda_{-} & \frac{1}{2}\omega_{5} & \frac{1}{2}\omega_{3} & \frac{1}{6}\lambda_{\eta} & \frac{1}{2}\lambda_{+} & \frac{1}{2}\omega_{5} & \frac{1}{2}\omega_{3} & \frac{1}{2}\lambda_{\eta} & \frac{1}{\sqrt{2}}\lambda_{3} & \frac{1}{3\sqrt{2}}\lambda_{\eta} & \frac{1}{\sqrt{2}}\omega_{3}\\
\frac{1}{3\sqrt{2}}\lambda_{H} & \frac{1}{\sqrt{2}}\omega_{2} & \frac{1}{\sqrt{2}}\omega_{1} & \frac{1}{\sqrt{2}}\lambda_{3} & \frac{1}{3\sqrt{2}}\lambda_{H} & \frac{1}{\sqrt{2}}\omega_{2} & \frac{1}{\sqrt{2}}\omega_{1} & \frac{1}{\sqrt{2}}\lambda_{3} & \frac{2}{3}\lambda_{H} & \lambda_{34} & \omega_{1}\\
\frac{1}{\sqrt{2}}\lambda_{3} & \frac{1}{\sqrt{2}}\omega_{5} & \frac{1}{\sqrt{2}}\omega_{3} & \frac{1}{3\sqrt{2}}\lambda_{\eta} & \frac{1}{\sqrt{2}}\lambda_{3} & \frac{1}{\sqrt{2}}\omega_{5} & \frac{1}{\sqrt{2}}\omega_{3} & \frac{1}{3\sqrt{2}}\lambda_{\eta} & \lambda_{34} & \frac{2}{3}\lambda_{\eta} & \omega_{3}\\
\frac{1}{\sqrt{2}}\omega_{1} & \frac{1}{\sqrt{2}}\omega_{4} & \frac{1}{3}\lambda_{\Phi} & \frac{1}{\sqrt{2}}\omega_{3} & \frac{1}{\sqrt{2}}\omega_{1} & \frac{1}{\sqrt{2}}\omega_{4} & \frac{1}{3\sqrt{2}}\lambda_{\Phi} & \frac{1}{\sqrt{2}}\omega_{3} & \omega_{1} & \omega_{3} & \frac{2}{3}\lambda_{\Phi}
\end{array}\right),
\end{equation}
where $\lambda_{34}\equiv\lambda_{3}+\lambda_{4}$. The eigenvalues
of this matrix are determined numerically. The second submatrix $\mathcal{M}^{N}_{2}$
defined in the basis 
\begin{eqnarray*}
\{h_{1}h_{2},h_{1}h_{3},h_{2}h_{3},h_{2}h_{4},h_{3}h_{4},a_{1}a_{2},a_{1}a_{3},a_{2}a_{3},a_{2}a_{4},a_{3}a_{4},h_{1}a_{2},\\
h_{1}a_{3},h_{2}a_{1},h_{2}a_{2},h_{2}a_{3},h_{2}a_{4},h_{3}a_{1},h_{3}a_{2},h_{3}a_{3},h_{3}a_{4},h_{4}a_{2},h_{4}a_{3}\}
\end{eqnarray*}
yields a diagonal matrix with degenerate eigenvalues given by $\omega_{1},\omega_{2},\omega_{3},\omega_{4},\omega_{5},\frac{\lambda_{s}}{3}$
and $\frac{\lambda_{\Phi}}{3}$ . 
The third submatrix $\mathcal{M}^{N}_{3}$ corresponds to the basis
$\left\{ h_{4}a_{4},h_{1}h_{4},a_{1}a_{4},h_{1}a_{1},h_{1}a_{4},h_{4}a_{1},h^{+}_{1}h^{-}_{2},h^{+}_{2}h^{-}_{1}\right\} $
and is given by 
\begin{align}
\mathcal{M}^{N}_{3} & =\left(\begin{array}{cccccccc}
\frac{\lambda_{\eta}}{3} & 0 & 0 & \lambda_{5} & 0 & 0 & 0 & 0\\
0 & \lambda_{+} & \lambda_{5} & 0 & 0 & 0 & \frac{1}{2}\lambda_{45} & \frac{1}{2}\lambda_{45}\\
0 & \lambda_{5} & \lambda_{+} & 0 & 0 & 0 & \frac{1}{2}\lambda_{45} & \frac{1}{2}\lambda_{45}\\
\lambda_{5} & 0 & 0 & \frac{\lambda_{H}}{3} & 0 & 0 & 0 & 0\\
0 & 0 & 0 & 0 & \lambda_{-} & \lambda_{5} & -\frac{i}{2}\lambda^{*}_{45} & \frac{i}{2}\lambda^{*}_{45}\\
0 & 0 & 0 & 0 & \lambda_{5} & \lambda_{-} & \frac{i}{2}\lambda^{*}_{45} & -\frac{i}{2}\lambda^{*}_{45}\\
0 & \frac{1}{2}\lambda_{45} & \frac{1}{2}\lambda_{45} & 0 & \frac{i}{2}\lambda^{*}_{45} & -\frac{i}{2}\lambda^{*}_{45} & \lambda_{34} & 2\lambda_{5}\\
0 & \frac{1}{2}\lambda_{45} & \frac{1}{2}\lambda_{45} & 0 & -\frac{i}{2}\lambda^{*}_{45} & \frac{i}{2}\lambda^{*}_{45} & 2\lambda_{5} & \lambda_{34}
\end{array}\right),
\end{align}

with $\lambda_{45}=\lambda_{4}+\lambda_{5}$ and $\lambda^{*}_{45}=\lambda_{4}-\lambda_{5}$.
This leads to the eigenvalues $\lambda_{3}\pm\lambda_{5},\,\lambda_{34},\lambda_{3}+2\lambda_{4}-3\lambda_{5}$
and $\frac{1}{6}(\lambda_{H}+\lambda_{\eta}\pm\sqrt{(\lambda_{\eta}-\lambda_{H})^{2}+36\lambda^{2}_{5}})$.

The scalar scattering matrix of the charged states is described by
a $16\times16$ matrix defined in the basis $\left\{ h_{1}h^{+}_{1},h_{2}h^{+}_{2},a_{1}h^{+}_{1},a_{2}h^{+}_{2},h_{1}h^{+}_{2},h_{3}h^{+}_{2},h_{4}h^{+}_{2},h_{2}h^{+}_{1},h_{3}h^{+}_{1},h_{4}h^{+}_{1},a_{1}h^{+}_{2},a_{3}h^{+}_{2},a_{4}h^{+}_{2},a_{2}h^{+}_{1},a_{3}h^{+}_{1},a_{4}h^{+}_{1}\right\} $.
This matrix can be decomposed as 
\begin{equation}
\mathcal{M}^{ch}_{16\times16}=\left(\begin{array}{cc}
\left(\mathcal{M}^{ch}_{1}\right)_{8\times8} & 0\\
0 & \left(\mathcal{M}^{ch}_{2}\right)_{8\times8}
\end{array}\right),
\end{equation}

with the first matrix defined in the basis $\left\{ h_{2}h^{+}_{2},a_{2}h^{+}_{2},h_{3}h^{+}_{2},h_{2}h^{+}_{1},h_{3}h^{+}_{1},a_{3}h^{+}_{2},a_{2}h^{+}_{1},a_{3}h^{+}_{1}\right\} $
by a diagonal matrix with the degenerate eigenvalues $\omega_{1},\omega_{2},\omega_{3}$
and $\omega_{5}$. 
The second submatrix is defined in the basis $\left\{ h_{1}h^{+}_{1},a_{1}h^{+}_{1},h_{1}h^{+}_{2},h_{4}h^{+}_{2},h_{4}h^{+}_{1},a_{1}h^{+}_{2},a_{4}h^{+}_{2},a_{4}h^{+}_{1}\right\} $
as 
\begin{align}
\mathcal{M}^{ch}_{2} & =\left(\begin{array}{cccccccc}
\frac{1}{3}\lambda_{H} & 0 & 0 & \frac{1}{2}\lambda_{45} & 0 & 0 & \frac{i}{2}\lambda^{*}_{45} & 0\\
0 & \frac{1}{3}\lambda_{H} & 0 & -\frac{i}{2}\lambda^{*}_{45} & 0 & 0 & \frac{1}{2}\lambda_{45} & 0\\
0 & 0 & \lambda_{3} & 0 & \frac{1}{2}\lambda_{45} & 0 & 0 & -\frac{i}{2}\lambda^{*}_{45}\\
\frac{1}{2}\lambda_{45} & \frac{i}{2}\lambda^{*}_{45} & 0 & \frac{1}{3}\lambda_{\eta} & 0 & 0 & 0 & 0\\
0 & 0 & \frac{1}{2}\lambda_{45} & 0 & \lambda_{3} & -\frac{i}{2}\lambda^{*}_{45} & 0 & 0\\
0 & 0 & 0 & 0 & \frac{i}{2}\lambda^{*}_{45} & \lambda_{3} & 0 & \frac{1}{2}\lambda_{45}\\
-\frac{i}{2}\lambda^{*}_{45} & \frac{1}{2}\lambda_{45} & 0 & 0 & 0 & 0 & \frac{1}{3}\lambda_{\eta} & 0\\
0 & 0 & \frac{i}{2}\lambda^{*}_{45} & 0 & 0 & \frac{1}{2}\lambda_{45} & 0 & \lambda_{3}
\end{array}\right),
\end{align}
\end{widetext}

that gives the eigenvalues $\lambda_{3}\pm\lambda_{4},\,\lambda_{3}\pm\lambda_{5},\,\frac{1}{6}(\lambda_{\eta}+\lambda_{H}\pm\sqrt{(\lambda_{\eta}-\lambda_{H})^{2}+36\lambda^{2}_{4}})$
and $\frac{1}{6}(\lambda_{H}+\lambda_{\eta}\pm\sqrt{(\lambda_{\eta}-\lambda_{H})^{2}+36\lambda^{2}_{5}})$.

\section{Scalar Couplings~\label{app:SV}}

The trilinear couplings among the neutral scalar fields are derived
from the scalar potential in Eq.~\eqref{eq:V} via the third-order
partial derivatives
\begin{equation}
\lambda_{H_{i}H_{j}H_{k}}=\left.\frac{\partial^{3}V}{\partial H_{i}\partial H_{j}\partial H_{k}}\right|_{\langle h_{i}\rangle=v_{i}},
\end{equation}
with $H_i=H_{0,1,2,3},A_0$.
Evaluating this expression yields: 
\begin{widetext}
\begin{alignat}{1}
\lambda_{H_{1}H_{1}H_{1}} & =3\omega_{1}R_{11}R_{13}\bigl(\upsilon_{\Phi}R_{11}+\upsilon R_{13}\bigr)+3\omega_{2}R_{11}R_{12}\bigl(\upsilon_{s}R_{11}+\upsilon R_{12}\bigr)\nonumber \\
 & \quad+3\omega_{4}R_{12}R_{13}\bigl(\upsilon_{\Phi}R_{12}+\upsilon_{s}R_{13}\bigr)+\lambda_{H}\upsilon R^{3}_{11}+\lambda_{s}\upsilon_{s}R^{3}_{12}+\lambda_{\Phi}\upsilon_{\Phi}R^{3}_{13},\nonumber \\
\lambda_{H_{2}H_{2}H_{2}} & =3\omega_{1}R_{21}R_{23}\bigl(\upsilon_{\Phi}R_{21}+\upsilon R_{23}\bigr)+3\omega_{2}R_{21}R_{22}\bigl(\upsilon_{s}R_{21}+\upsilon R_{22}\bigr)\nonumber \\
 & \quad+3\omega_{4}R_{22}R_{23}\bigl(\upsilon_{s}R_{23}+\upsilon_{\Phi}R_{22}\bigr)+\lambda_{H}\upsilon R^{3}_{21}+\lambda_{s}\upsilon_{s}R^{3}_{22}+\lambda_{\Phi}\upsilon_{\Phi}R^{3}_{23},\nonumber \\
\lambda_{H_{3}H_{3}H_{3}} & =3\omega_{1}R_{31}R_{33}\bigl(\upsilon_{\Phi}R_{31}+\upsilon R_{33}\bigr)\quad+3\omega_{2}R_{31}R_{32}\bigl(\upsilon_{s}R_{31}+\upsilon R_{32}\bigr)\nonumber \\
 & \quad+3\omega_{4}R_{33}R_{32}\bigl(\upsilon_{s}R_{33}+\upsilon_{\Phi}R_{32}\bigr)\quad+\lambda_{H}\upsilon R^{3}_{31}+\lambda_{s}\upsilon_{s}R^{3}_{32}+\lambda_{\Phi}\upsilon_{\Phi}R^{3}_{33} \\
\lambda_{H_{i}H_{0}H_{0}} & =\lambda_{+}\upsilon R_{i1}+\omega_{5}\upsilon_{s}R_{i2}+\omega_{3}\upsilon_{\Phi}R_{i3},\,\,\,\,\lambda_{H_{i}H^{+}H^{-}}=\lambda_{3}\upsilon R_{i1}+\omega_{5}\upsilon_{s}R_{i2}+\omega_{3}\upsilon_{\Phi}R_{i3}\nonumber \\
\lambda_{H_{i}A_{0}A_{0}} & =\lambda_{-}\upsilon R_{i1}+\omega_{5}\upsilon_{s}R_{i2}+\omega_{3}\upsilon_{\Phi}R_{i3},\,\,\,\lambda_{H_{0}H_{i}H_{j}}=\lambda_{H_{0}H^{+}H^{-}}=\lambda_{H_{0}A_{0}A_{0}}=\lambda_{H_{0}H_{0}A_{0}}=0.
\end{alignat}

Due to the $Z_{2}$ symmetry, all vertices $H_{0}VV'$ and $H_{0}H_{i}VV'$
(with $i\ne0$) vanish, i.e., $\lambda^{\mu\nu}_{H_{0}VV'}=\lambda^{\mu\nu}_{H_{0}H_{i}VV'}=0$.
In addition, all mixed photon and $Z$ type quartic couplings vanish
$\lambda^{\mu\nu}_{H_{i}\gamma Z}=\lambda^{\mu\nu}_{H_{i}\gamma Z^{\prime}}=\lambda^{\mu\nu}_{H_{i}\gamma Z^{\prime\prime}}=\lambda^{\mu\nu}_{H_{i}ZZ^{\prime}}=\lambda^{\mu\nu}_{H_{i}ZZ^{\prime\prime}}=0$.
with $i=1,2,3$. The non-vanishing trilinear scalar gauge couplings
involving neutral scalars are expressed as:
\begin{align}
\lambda^{\mu\nu}_{H_{i}ZZ} & =\eta^{\mu\nu}\frac{\upsilon\,e^{2}R_{i1}}{2s^{2}_{W}c^{2}_{W}},\,\,\,\lambda^{\mu\nu}_{H_{i}Z^{\prime}Z^{\prime}}=\eta^{\mu\nu}\left[\frac{(g_{H}c_{\alpha}c_{W}+es_{\alpha})^{2}}{2c^{2}_{W}}R_{i3}\upsilon_{\Phi}+\frac{8e^{2}s^{2}_{\alpha}}{c^{2}_{W}}R_{i2}\upsilon_{s}\right],\nonumber \\
\lambda^{\mu\nu}_{H_{i}Z^{\prime\prime}Z^{\prime\prime}} & =\eta^{\mu\nu}\left[\frac{(-g_{H}s_{\alpha}c_{W}+ec_{\alpha})^{2}}{2c^{2}_{W}}R_{i3}\upsilon_{\Phi}+\frac{8e^{2}c^{2}_{\alpha}}{c^{2}_{W}}R_{i2}\upsilon_{s}\right],\nonumber \\
\lambda^{\mu\nu}_{H_{i}Z^{\prime}Z^{\prime\prime}} & =\eta^{\mu\nu}\left[-\frac{1}{2c^{2}_{W}}\left(-g_{H}s_{\alpha}c_{W}+ec_{\alpha}\right)\left(g_{H}c_{\alpha}c_{W}+es_{\alpha}\right)R_{i3}\upsilon_{\Phi}-\frac{8e^{2}s_{\alpha}c_{\alpha}}{c^{2}_{W}}R_{i2}\upsilon_{s}\right]\nonumber\\
\lambda^{\mu\nu}_{Q_{0}\bar{Q}_{0}Z^{\prime}} & = \eta^{\mu\nu} g_{H}c_{\alpha},\qquad~\lambda^{\mu\nu}_{Q_{0}\bar{Q}_{0}Z^{\prime\prime}}=\eta^{\mu\nu} g_{H}s_{\alpha},
\end{align}
while 
\begin{equation}
\lambda^{\mu\nu}_{H^{+}Q_{0}Z}=\lambda^{\mu\nu}_{H^{+}Q_{0}Z^{\prime}}=\lambda^{\mu\nu}_{H^{+}Q_{0}Z^{\prime\prime}}=\lambda^{\mu\nu}_{Q_{0}\bar{Q}_{0}\gamma}=\lambda^{\mu\nu}_{Q_{0}\bar{Q}_{0}Z}=0.
\end{equation}
\end{widetext}

The neutral scalar couplings to $Q_{0}\bar{Q}_{0}$ are given by 
\begin{equation}
\lambda^{\mu\nu}_{H_{i}Q_{0}\bar{Q}_{0}}=\eta^{\mu\nu}\frac{1}{2}\,\upsilon_{\Phi}\,R_{i3}\,g^{2}_{H},\qquad(i=1,2,3),
\end{equation}
while $\lambda^{\mu\nu}_{H_{0}W'_{1}W'_{2}}=0$. In our model, the
trilinear couplings involving charged scalars are vanishing $\lambda^{\mu\nu}_{H^{+}W^{-}Z}=\lambda^{\mu\nu}_{H^{+}W^{-}Z^{\prime}}=\lambda^{\mu\nu}_{H^{+}W^{-}Z^{\prime\prime}}=0$.
For $W$ gauge boson, we have 
\begin{equation}
\lambda^{\mu\nu}_{H_{i}W^{+}W^{-}}=\eta^{\mu\nu}\frac{\upsilon\,e^{2}R_{i1}}{2s^{2}_{W}}.
\end{equation}

Regarding the quartic gauge-scalar couplings, all mixed photon--$Z$-type
quartic couplings vanish, 
\begin{widetext}
\begin{equation}
\lambda^{\mu\nu}_{H_{i}H_{j}\gamma Z}=\lambda^{\mu\nu}_{H_{i}H_{j}\gamma Z^{\prime}}=\lambda^{\mu\nu}_{H_{i}H_{j}\gamma Z^{\prime\prime}}=\lambda^{\mu\nu}_{H_{i}H_{j}ZZ^{\prime}}=\lambda^{\mu\nu}_{H_{i}H_{j}ZZ^{\prime\prime}}=0,
\end{equation}
with $i,j=1,2,3$. The non-zero quartic couplings are given by 
\begin{align}
\lambda^{\mu\nu}_{H_{i}H_{j}ZZ} & =\eta^{\mu\nu}\frac{e^{2}R_{i1}R_{j1}}{2s^{2}_{W}c^{2}_{W}},\,\,\,\lambda^{\mu\nu}_{H_{i}H_{j}Z^{\prime}Z^{\prime}}=\eta^{\mu\nu}\left[\frac{\left(g_{H}c_{\alpha}c_{W}+es_{\alpha}\right)^{2}}{2c^{2}_{W}}R_{i3}R_{j3}+\frac{8e^{2}s^{2}_{\alpha}}{c^{2}_{W}}R_{i2}R_{j2}\right],\nonumber \\
\lambda^{\mu\nu}_{H_{i}H_{j}Z^{\prime\prime}Z^{\prime\prime}} & =\eta^{\mu\nu}\left[\frac{\left(g_{H}s_{\alpha}c_{W}-ec_{\alpha}\right)^{2}}{2c^{2}_{W}}R_{i3}R_{j3}+\frac{8e^{2}c^{2}_{\alpha}}{c^{2}_{W}}R_{i2}R_{j2}\right],\nonumber \\
\lambda^{\mu\nu}_{H_{i}H_{j}Z^{\prime}Z^{\prime\prime}} & =\eta^{\mu\nu}\Big[\frac{1}{2c^{2}_{W}}\left(g_{H}s_{\alpha}c_{W}-ec_{\alpha}\right)\left(g_{H}c_{\alpha}c_{W}+es_{\alpha}\right)R_{i3}R_{j3}-\frac{8e^{2}s_{\alpha}c_{\alpha}}{c^{2}_{W}}R_{i2}R_{j2}\Big],
\end{align}
and 
\begin{equation}
\lambda^{\mu\nu}_{H_{i}H_{j}W^{+}W^{-}}=\eta^{\mu\nu}\frac{e^{2}}{2s^{2}_{W}}R_{i1}R_{j1},\qquad\lambda^{\mu\nu}_{H_{0}H_{0}W^{+}W^{-}}=\eta^{\mu\nu}\frac{e^{2}}{2s^{2}_{W}}.
\end{equation}
\end{widetext}

The non-zero $Q_{0}\bar{Q}_{0}$ couplings are 
\begin{equation}
\lambda^{\mu\nu}_{H_{i}H_{j}Q_{0}\bar{Q}_{0}}=\eta^{\mu\nu}\frac{g^{2}_{H}}{2}R_{i3}R_{j3}.
\end{equation}
Furthermore, we have 
\begin{align}
\lambda^{\mu\nu}_{H_{0}H_{0}ZZ}= & \eta^{\mu\nu}\frac{e^{2}}{2s^{2}_{W}c^{2}_{W}}
\end{align}
Finally, the following couplings vanish: 
\begin{align}
 & \lambda^{\mu\nu}_{H_{0}H_{0}AZ}=\lambda^{\mu\nu}_{H_{0}H_{0}AZ^{\prime}}=\lambda^{\mu\nu}_{H_{0}H_{0}AZ^{\prime\prime}}=\nonumber \\
 & \lambda^{\mu\nu}_{H_{0}H_{0}AA}=\lambda^{\mu\nu}_{H_{0}H_{0}Z^{\prime}Z^{\prime}}=\lambda^{\mu\nu}_{H_{0}H_{0}Z^{\prime\prime}Z^{\prime\prime}}=\nonumber \\
 & \lambda^{\mu\nu}_{H_{0}H_{0}Z^{\prime}Z^{\prime\prime}}=\lambda^{\mu\nu}_{H_{0}H_{0}ZZ^{\prime\prime}}=\lambda^{\mu\nu}_{H_{0}H_{0}ZZ^{\prime}}=0.
\end{align}

\section{Coupling coefficients for $Z^{\prime}$ and $Z^{\prime\prime}$~\label{app:kaapa}}

The squared effective couplings are flavor-diagonal
and universal across generations. Explicitly, the couplings for the
$Z^{\prime}$ gauge boson are given by

\begin{align}
\kappa^{2}_{Z^{\prime},\ell} & =\frac{16s^{2}_{\alpha}s^{2}_{W}}{(4s^{2}_{W}-1)^{2}+1}\,,\qquad(\ell=e,\mu,\tau)\\[1ex]
\kappa^{2}_{Z^{\prime},u} & =\frac{16s^{2}_{\alpha}s^{2}_{W}}{9\left[\left(1-\frac{8}{3}s^{2}_{W}\right)^{2}+1\right]}\,,\qquad(u=u,c,t)\\[1ex]
\kappa^{2}_{Z^{\prime},d} & =\frac{16s^{2}_{\alpha}s^{2}_{W}}{9\left[\left(-1+\frac{4}{3}s^{2}_{W}\right)^{2}+1\right]}\,,\qquad(d=d,s,b)\\[1ex]
\kappa^{2}_{Z^{\prime},\nu} & ={4s^{2}_{\alpha}s^{2}_{W}}\,,\qquad(\nu=\nu_{e},\nu_{\mu},\nu_{\tau})\,,
\end{align}
whereas the corresponding couplings for the $Z^{\prime\prime}$ boson are obtained
by replacing $s^{2}_{\alpha}\to c^{2}_{\alpha}$:

\begin{align}
\kappa^{2}_{Z^{\prime\prime},\ell} & =\frac{16c^{2}_{\alpha}s^{2}_{W}}{(4s^{2}_{W}-1)^{2}+1}\,,\qquad(\ell=e,\mu,\tau)\\[1ex]
\kappa^{2}_{Z^{\prime\prime},u} & =\frac{16c^{2}_{\alpha}s^{2}_{W}}{9\left[\left(1-\frac{8}{3}s^{2}_{W}\right)^{2}+1\right]}\,,\qquad(u=u,c,t)\\[1ex]
\kappa^{2}_{Z^{\prime\prime},d} & =\frac{16c^{2}_{\alpha}s^{2}_{W}}{9\left[\left(-1+\frac{4}{3}s^{2}_{W}\right)^{2}+1\right]}\,,\qquad(d=d,s,b)\\[1ex]
\kappa^{2}_{Z^{\prime\prime},\nu} & ={4c^{2}_{\alpha}s^{2}_{W}}\,,\qquad(\nu=\nu_{e},\nu_{\mu},\nu_{\tau})\,.
\end{align}

 \bibliographystyle{utphys}
\bibliography{references}

\end{document}